\documentclass[journal]{IEEEtran}

\usepackage{color}
\usepackage{bbm}
\usepackage[super]{nth}
\usepackage{amsthm}
\usepackage{algorithm}
\usepackage{amssymb}
\usepackage{multirow}
\usepackage[pdftex]{graphicx}
\usepackage{epstopdf}
\usepackage{caption}
\usepackage{cuted}
\usepackage{float}
\usepackage{stfloats}

\usepackage{tabularx}  
\usepackage{arydshln,leftidx,mathtools} 
\newcounter{mytempeqncnt}

\theoremstyle{definition}
\newtheorem{theorem}{Theorem}
\newtheorem{proposition}{Proposition}

\newtheorem{lemma}{Lemma}

\usepackage{cite}

\ifCLASSINFOpdf
 \usepackage[pdftex]{graphicx}

\else
\fi
\usepackage{amsmath}
\usepackage{algorithmic}

\usepackage{array}

\usepackage{subfig}

\begin{document}
\title{Coded Distributed Computing  over \\ Packet Erasure Channels}
%
%
%


\author{
Dong-Jun~Han,~\IEEEmembership{Student Member,~IEEE,}
Jy-yong Sohn,~\IEEEmembership{Student Member,~IEEE,}
Jaekyun~Moon,~\IEEEmembership{Fellow,~IEEE}  

\thanks{The authors are with the School of Electrical Engineering, Korea Advanced Institute of Science and Technology (KAIST), Daejoen 34141, South Korea (e-mail: djhan93@kaist.ac.kr; jysohn1108@kaist.ac.kr; jmoon@kaist.edu).
}
}
\maketitle

\begin{abstract}
Coded computation is a framework which provides redundancy in distributed computing systems to speed up large-scale tasks. Although most existing works assume an error-free scenarios in a master-worker setup, the link failures are common in current wired/wireless networks. In this paper, we consider the straggler problem in coded distributed computing with link failures, by modeling the links between the master node and worker nodes as packet erasure channels. When the master fails to detect the received signal, retransmission is required for each worker which increases the overall run-time to finish the task. We first investigate the expected overall run-time in this setting using an $(n,k)$ maximum distance separable (MDS) code. We obtain the lower and upper bounds on the latency in closed-forms and give guidelines to design MDS code depending on the packet erasure probability. Finally, we consider a setup where the number of retransmissions is limited due to the bandwidth constraint. By formulating practical optimization problems related to latency, transmission bandwidth and probability of successful computation, we obtain achievable performance curves as a function of packet erasure probability.

\end{abstract}
\begin{IEEEkeywords}
Distributed computing, Packet erasure channels
\end{IEEEkeywords}

%
\IEEEpeerreviewmaketitle

\section{Introduction}
%
%
%
%

\IEEEPARstart{S}{olving} massive-scale computational tasks for machine learning algorithms and data analytics is one of the most rewarding challenges in the era of big data. Modern large-scale computational tasks cannot be solved in a single machine (or worker) anymore. Enabling large-scale computations, distributed computing systems such as MapReduce \cite{mapreduce}, Apache Spark \cite{Spark} and Amazon EC2 have received significant attention in recent years. In distributed computing systems, the large-scale computational tasks are divided into several subtasks, which are computed by different workers in parallel. This parallelization helps to reduce the overall run-time to finish the task and thus enables to handle large-scale computing tasks. However, it is observed that workers that are significantly slow than the average, called \textit{stragglers}, critically slow down the computing time in practical distributed computing systems.  


To alleviate the effect of stragglers, the authors of \cite{KLee1} proposed a new framework called \textit{coded computation}, which uses error correction codes to provide redundancy in distributed matrix multiplications. By applying maximum distance separable (MDS) codes, it was shown that total computation time of the coded scheme can achieve an order-wise improvement over uncoded scheme. More recently, coded computation has been applied to various other distributed computing scenarios. In \cite{High}, the authors proposed the use of product codes for high-dimensional matrix multiplication. The authors of \cite{Pcode} also targets high-dimensional matrix multiplication by proposing the use of polynomial codes. The use of short dot product for computing large linear transforms is proposed in \cite{shortdot}, and Gradient coding is proposed with applications to distributed gradient descent in \cite{Gradient1, Gradient2, Gradient3}. Codes for convolution and Fourier transform are studied in \cite{convolution} and \cite{Fourier}, respectively. To compensate the drawback of previous studies which completely ignore the work done by stragglers, the authors of \cite{Draper1, Draper2} introduce the idea of exploiting the work completed by the stragglers.

Mitigating the effect of stragglers over more practical settings are also being studied in recent years. The authors of \cite{heterogeneous} propose an algorithm for speeding up distributed matrix multiplication over heterogeneous clusters, where a variety of computing machines coexist with different capabilities. In \cite{HPark}, a hierarchical coding scheme is proposed to combat stragglers in practical distributed computing systems with multiple racks. The authors of \cite{wireless} also consider a setting where the workers
are connected wirelessly to the master.

Codes are also introduced in various studies targeting data shuffling applications \cite{SLi1,SLi2,MAttia,SLi3,wireless_datasuffling,shuffling_limit} in MapReduce style setups. In \cite{SLi3}, a unifed coding framework is proposed for distributed computing, and trade-off between latency of computation and load of communication is studied. A scalable framework is studied in \cite{wireless_datasuffling} for minimizing the communication bandwidth in wireless distributed computing scenario. In \cite{shuffling_limit}, fundamental limits of data shuffling is investigated for distributed learning. 

\subsection{Motivation}
In most of the existing studies, it is assumed that the links between the master node and worker nodes are error-free. However, link failures and device failures are common in current wired/wireless networks. It is reported in \cite{Error_datacenter} that the transmitted packets
often get lost in current data centers by various factors such
as congestion or load balancer issues in current data centers. In wireless networks, the transmitted packets can be lost by channel fading or interference. Based on these observations, the authors of \cite{Erasure_storage} considered link failures in the context of distributed storage systems. In this paper, we consider link failures for modern  distributed computing systems. The transmission failure
at a specific worker necessitates packet retransmission, which
would in turn increase the overall run-time for the given task. Moreover, when
the bandwidth of the system is limited, the master would not
be able to collect timely results from the workers due to link
failures. A reliable solution to deal with these problems is
required in practical distributed computing systems.

\subsection{Contribution}
In this paper, we model the links between the master node and worker nodes as packet erasure channels, which is inspired by the link failures in practical distributed computing systems. We consider straggler problems in matrix-vector multiplication which is a key computation in many machine learning algorithms. We first separate the run-time at each worker into computation time and communication time. This approach is also taken in the previous work \cite{wireless} on coded computation over wireless networks, under the error-free assumption. Compared to the setting in \cite{wireless}, in our work, packet loss and retransmissions are considered. Thus, the communication time at each worker becomes a random variable which depends on the packet erasure probability. 

We analyze the latency in this setting using an $(n,k)$ MDS code. By considering the average computation/communication time and erasure probability, we obtain the overall run-time of the task by analyzing the continuous-time Markov chain for maximum $k$ (when each node computes a single inner product). We also find the lower and upper bounds on the latency in closed-forms. Based on the closed-form expressions we give insights on the bounds and show that the overall run-time becomes log$(n)$ times faster than the uncoded scheme by using $(n,k)$ MDS code. Then for general $k$, we give guidelines to design $(n,k)$ MDS code depending on the erasure probability $\epsilon$. Finally, we consider a setup where the number of retransmissions is limited due to the bandwidth constraint. By formulating practical optimization problems related to latency, transmission bandwidth and probability of successful computation, we obtain achievable performance curves as a function of packet erasure probability.

\subsection{Organization and Notations}
The rest of this paper is organized as follows. In Section \ref{sec:problem_statement}, we describe the master-worker setup with link failures and define a problem statement. Latency analysis for maximum $k$ is performed in Section \ref{sec:analysis1}, and the case for general $k$ is discussed in Section \ref{sec:analysis2}. Then in Section \ref{sec:Finite}, analysis is performed under the setting of limited number of retransmissions. Finally, concluding remarks are drawn in Section \ref{sec:conclusion}.

The definitions and notations used in this paper are summarized as follows. 
For $n$ random variables, the $k^{\text{th}}$ order statistic is denoted by the $k^{\text{th}}$ smallest one of $n$. We denote the set $\{1,2,...n\}$ by $[n]$ for $n\in\mathbb{N}$, where $\mathbb{N}$ is the set of positive integers. We also denote the $k^{\text{th}}$ order statistics of $n$ independent random variables $(X_1,X_2,...X_n)$ by $X_{(k)}$. We denote $f(n)=\Theta(g(n))$ if there exist $c_1,c_2,n_0>0$ such that $\forall n>n_0$, $c_1g(n)\leq f(n) \leq c_2g(n)$.


\section{Problem Statement}\label{sec:problem_statement}
Consider a master-worker setup in wired/wireless network with $n$ worker nodes connected to a single master node. For the wireless scenario, frequency division multiple access (FDMA) can be used so that the signals sent from the workers are detected simultaneously at the master without interference. We study a matrix-vector multiplication problem, where the goal is to compute the output $\mathbf{y}=\mathbf{A}\mathbf{x}$ for a given matrix $\mathbf{A}\in \mathbb{R}^{m \times d}$ and an input vector $\mathbf{x}\in \mathbb{R}^{d\times 1}$. The matrix $\mathbf{A}$ is divided into $k$ submatrices to obtain $\mathbf{A}_i\in \mathbb{R}^{\frac{m}{k} \times d}$, $i\in [k]$. Then, an $(n, k)$ MDS code is applied to construct $\tilde{\mathbf{A}}_i\in \mathbb{R}^{\frac{m}{k} \times d}$, $i\in [n]$, which are distributed across the $n$ worker nodes ($n>k$). For a given input vector $\mathbf{x}$, each worker $i$ computes $\tilde{\mathbf{A}}_i\mathbf{x}$ and sends the result to the master. By receiving $k$ out of $n$ results, the master node can recover the desired output $\mathbf{A}\mathbf{x}$. 

We assume that the required time at each worker to compute a single inner product of vectors of size $d$ follows an exponential distribution with rate $\mu_1$. Since $\frac{m}{k}$ inner products are to be computed each worker, similar to the models in \cite{wireless,heterogeneous}, we assume that $X_i$, the computation time of worker $i$ follows an exponential distribution with rate $\frac{k\mu_1}{m}$:
\begin{equation}
\text{Pr}[X_i\leq t]=1-e^{-\frac{k}{m}\mu_1t}. 
\end{equation}

After completing the given task, each worker transmits its results to the master through memoryless packet erasure channels where the packets are erased independently with probability $\epsilon$. For simplicity, we assume that each packet contains a single inner product. When a packet fails to be detected at the master, the corresponding packet is retransmitted. Otherwise, the worker transmits the next packet to the master. The work is finished when the master successfully detects the results of $k$ out of $n$ workers. The communication time for $j^{\text{th}}$ transmission of the $r^{\text{th}}$ inner product at each worker, $Y_{j,r}$, is also modeled as an exponential distribution (as the same as the communication time model in \cite{HPark}) with rate $\mu_2$. Then the average time for one packet transmission becomes $\frac{1}{\mu_2}$. Since each worker transmits $\frac{m}{k}$ inner products, the total communication time of worker $i$, denoted by $S_i$, is written as 
\begin{equation}
S_i=\sum_{r=1}^{\frac{m}{k}}\sum_{j=1}^{N_r}Y_{j,r} \label{eq:communication_time}
\end{equation}
where $\text{Pr}[Y_{j,r}\leq t]=1-e^{-\mu_2t}$ and $N_r$ is a geometric random variable with success probability $1-\epsilon$, i.e., $\text{Pr}[N_r=l]=(1-\epsilon)\epsilon^{l-1}$.

We define the overall run-time of the task as the sum of computation time and communication time to finish the task. Now the overall run-time using $(n, k)$ MDS code is written as 
\begin{equation}\label{eq:T}
T=\underset{i\in[n]}{k^{\text{th}}\text{min}}(X_i+S_i).
\end{equation}
We are interested in the expected overall run-time $\mathbb{E}[T]$.


\section{Latency Analysis for $k=m$}\label{sec:analysis1}
We start by reviewing some useful results on the order statistics. The expected value of $k^{\text{th}}$ order statistic of $n$ exponential random variables of rate $\mu$ is $\frac{H_n-H_{n-k}}{\mu}$, where $H_k=\sum_{i=1}^{k}\frac{1}{i}$. For $k=n$, the $k^{\text{th}}$ order statistic becomes $\frac{H_n}{\mu}$. The value $H_n$ can be approximated as $H_n\simeq \text{log}(n)+\eta$ for large $n$ where $\eta$ is a fixed constant.

In this section, we assume that each worker is capable of computing a single inner product, i.e., $k=m$. 
This is a realistic assumption especially when a massive number of Internet of Things (IoT) devices having small storage size coexist in the network for wireless distributed computing\cite{WDC}. The case for general $k$ will be discussed in Section \ref{sec:analysis2}. 

\subsection{Exact Overall Run-Time: Markov Chain Analysis}

We derive the overall run-time in terms $\mu_1$, $\mu_2$, $\epsilon$, $n$. We first state the following result on the communication time.
\begin{lemma}\label{lemma:sum_of_geometric_exponentials}
Assuming $k=m$, the random variable $S_i$ follows exponential distribution with rate $(1-\epsilon)\mu_2$, i.e., $\text{Pr}[S_i\leq t]=1-e^{-(1-\epsilon)\mu_2t}$.
\end{lemma}
\begin{IEEEproof}
See Appendix A.
\end{IEEEproof}

\textit{Remark} 1: By erasure probability $\epsilon$ and retransmission, Lemma 1 implies that the rate of communication time at each worker decreases by a factor of $1-\epsilon$ compared to the error-free scenario. That is, the expected communication time at each worker increases by a factor of $\frac{1}{1-\epsilon}$.

Based on Lemma 1, we obtain the following theorem which shows that $\mathbb{E}[T]$ can be obtained by analyzing the hitting time of a continuous-time Markov chain.
\begin{theorem}\label{thm:Markov}
Consider a continuous-time Markov chain $\mathbb{C}$ defined over 2-dimensional state space $(u,v)\in\{0, 1, ..., n\}\times\{0,  1,..., k \}$, where the transition rates are defined as follows.
\begin{itemize}
\item The transition rate from state $(u,v)$ to state $(u+1,v)$ is $(n-u)\mu_1$ if $v\leq u<n$, and $0$ otherwise.
\item The transition rate from state $(u,v)$ to state $(u,v+1)$ is $(u-v)(1-\epsilon)\mu_2$ if $0\leq v < \text{min}(u,k)$, and $0$ otherwise.
\end{itemize}
Then, the expected hitting time of $\mathbb{C}$ from state $(0, 0)$ to the set of states $(u,v)$ with $v=k$ becomes the expected overall run-time $\mathbb{E}[T]$.
\end{theorem}
\begin{IEEEproof}
For a given time $t$, define integers $u,v$ as
\begin{equation}
u=|\{j\in[n]:X_{j} \leq t\}|  \label{eq:def_u}
\end{equation} 
\begin{equation}
v=|\{i\in[n]:X_{i}+S_i \leq t\}| \label{eq:def_v}
\end{equation} 
Here, $u$ can be viewed as the number of workers that finished computation by time $t$. For $v$, it is the number of workers that successfully transmitted the computation result to the master by time $t$. From these definitions, $v$ can be rewritten as 
\begin{equation}
v=|\{i\in[u]: X_{i}+S_i \leq t \}|. \label{eq:def_v2}
\end{equation}

From $\mathbb{E}[T]=\mathbb{E}[\underset{i\in[n]}{k^{\text{th}}\text{min}}(X_i+S_i)]$, it can be seen that the expected overall run-time is equal to the expected arrival time from $(0,0)$ to any $(u,v)$ with $v=k$. Thus, we define the state of the Markov chain $(u,v)$, over space $\{0, 1, ..., n\}\times\{0,  1,..., k \}$. 

By defining $u,v$ as above, the transition rates can be described as follows. By (\ref{eq:def_u}), for a given $t$, there are $n-u$ number of integers $j\in [n]$ satisfying $X_{j}>t$ at state $(u,v)$. Since $X_j$ follows exponential distribution with rate $\mu_1$, state transistion from $(u,v)$ to $(u+1,v)$ occurs with rate $(n-u)\mu_1$. The transition from $(u,v)$ from $(u+1,v)$ is possible for $u\geq v$, by the definitions (\ref{eq:def_u}), (\ref{eq:def_v}).

Now we consider the transition rate from $(u,v)$ to $(u,v+1)$ for a fixed $t$. Among $u$ number of integers $i$ satisfying $X_i\leq t$, there are $v$ of them which satisfy $X_i+S_i\leq t$ at state $(u,v)$. Recall that $S_i$ follows exponential distribution with rate $(1-\epsilon)\mu_2$ from Lemma \ref{lemma:sum_of_geometric_exponentials}. Thus, state transition from $(u,v)$ to $(u,v+1)$ occurs with rate $(u-v)(1-\epsilon)\mu_2$. Note that $v\in\{0,1,...,\text{min}(u,k)-1\}$. The described Markov chain is identical to $\mathbb{C}$, which completes the proof.
\end{IEEEproof}

\begin{figure}[t]
\centering
     \includegraphics[width=0.49\textwidth, trim=-0.5cm 0 0 0]{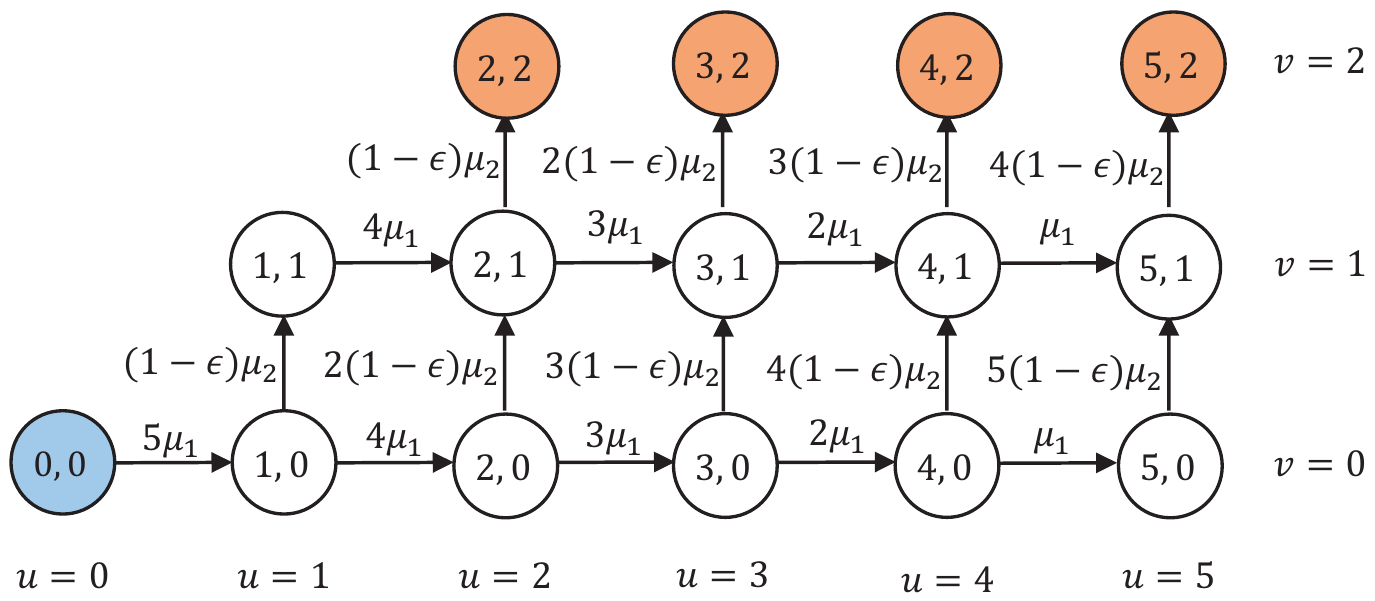}
\caption{State transition diagram example with $n=5$, $k=2$.}\label{fig:Markov_example}
\end{figure}

For an arbitrary state $(u,v)$, the first component $u$ can be viewed as the number of workers finished the computation, and the second component $v$ is the number of workers that successfully transmitted their result to the master. Fig. \ref{fig:Markov_example} shows an example of the transition state diagram assuming $n=5$, $k=2$. From the initial state $(0,0)$, the given task is finished when the Markov chain visits one of the states with $v=2$ for the first time. Since the computational results of a worker are
transmitted after the worker finishes the whole computation, $u\geq v$ always holds. The expected hitting time of the Markov chain can be computed performing the first-step analysis \cite{Markov_anlysis}.

\subsection{Lower and Upper Bounds}
While the exact $\mathbb{E}[T]$ is not derived in a closed-form, we provide the closed-form expressions of the lower and upper bounds of $\mathbb{E}[T]$ in the following theorem.
\begin{theorem}\label{thm:bounds}
Assuming $k=m$, the expected overall run-time $\mathbb{E}[T]$ is lower bounded as $\mathbb{E}[T]\geq \mathcal{L}$, where
\begin{equation}\label{eq:Lbound}
\mathcal{L}=\underset{i\in[k]}{\text{max}}(\frac{H_n-H_{n-k+i-1}}{\mu_1}+\frac{H_n-H_{n-i}}{(1-\epsilon)\mu_2}),
\end{equation}
which reduces to
\begin{equation}\label{eq:Lbound_reduced}
\mathcal{L}=\begin{dcases}
\frac{1}{n\mu_1}+\frac{H_n-H_{n-k}}{(1-\epsilon)\mu_2}, \ \ \ \ \ \ \ \ \ \ \  \epsilon\geq 1 - \frac{\mu_1}{\mu_2} \\
\frac{H_n-H_{n-k}}{\mu_1}+\frac{1}{n(1-\epsilon)\mu_2}, \ \ \ otherwise.
\end{dcases}
\end{equation}
Moreover, the expected overall run-time $\mathbb{E}[T]$ is upper bounded as $\mathbb{E}[T]\leq\mathcal{U}$, where
\begin{equation}\label{eq:Ubound}
\mathcal{U}= \underset{i\in[n]\setminus[k-1]}{\text{min}}(\frac{H_n-H_{i-k}}{\mu_1}+\frac{H_n-H_{n-i}}{(1-\epsilon)\mu_2}).
\end{equation}
\end{theorem}
\begin{IEEEproof}
See Appendix C. 
\end{IEEEproof}

From (\ref{eq:Lbound}) and (\ref{eq:Ubound}), it can be seen that the lower and upper bounds of $\mathbb{E}[T]$ are decreasing functions of $\epsilon$. Note that $\epsilon$ only decreases the terms of communication time (not computation time) by a factor of $\frac{1}{1-\epsilon}$ as in (\ref{eq:Lbound}) and (\ref{eq:Ubound}). The lower bound in (\ref{eq:Lbound}) can be rewritten as (\ref{eq:Lbound_reduced}) with two regimes classifed by equation $\epsilon= 1-\frac{\mu_1}{\mu_2}$. This equation is equivalent to 
\begin{equation}
\frac{1}{\mu_1}=\frac{1}{(1-\epsilon)\mu_2}
\end{equation}
which means that the average computation time is equal to the average communication time with retransmission.

For $\epsilon\geq 1-\frac{\mu_1}{\mu_2}$, the first term of the lower bound in (\ref{eq:Lbound_reduced}) is the expected value of minimum computation time among $n$ workers, while the second term is the expected value of the $k^{\text{th}}$ smallest communication time among $n$ workers. This is the case where all the workers finish the computation by time $\frac{1}{n\mu_1}$ and then start communication simultaneously, which can be viewed as an obvious lower bound. We show in Appendix C that the derived lower bound (\ref{eq:Lbound_reduced}) is the tightest among $k$ candidates of the bounds we obtained in (\ref{eq:Lbound}). For the upper bound, by defining $H_0=0$, we have $\mathcal{U}=\frac{H_n}{\mu_1}+\frac{H_n-H_{n-k}}{(1-\epsilon)\mu_2}$ by inserting $i=k$ at (\ref{eq:Ubound}). The first term $\frac{H_n}{\mu_1}$ is the expected value of maximum computation time among $n$ workers and the second term is the expected value of the $k^{\text{th}}$ smallest communication time among $n$ workers. Again, this is an obvious upper bound as it can be interpreted as the case where all $n$ workers starts communication simultaneously right after the slowest worker finishes its computation. The derived upper bound includes this kind of scenario. 

\begin{figure}
	\centering
	\subfloat[][$\mu_1=10$, $\mu_2=1$.]{\includegraphics[height=53mm ]{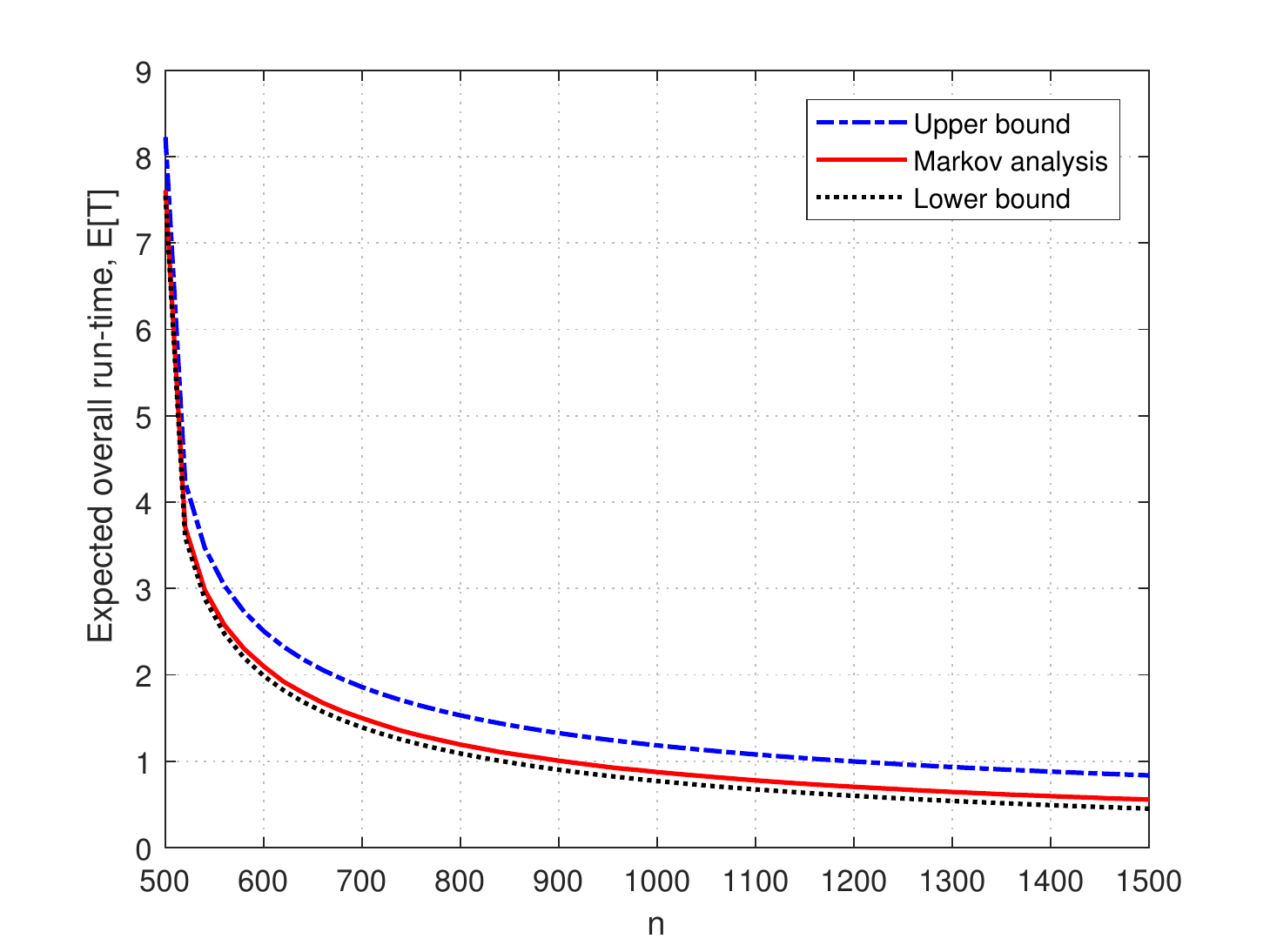}}
	\quad \quad
	\subfloat[][$\mu_1=1$, $\mu_2=10$.]{\includegraphics[height=53mm]{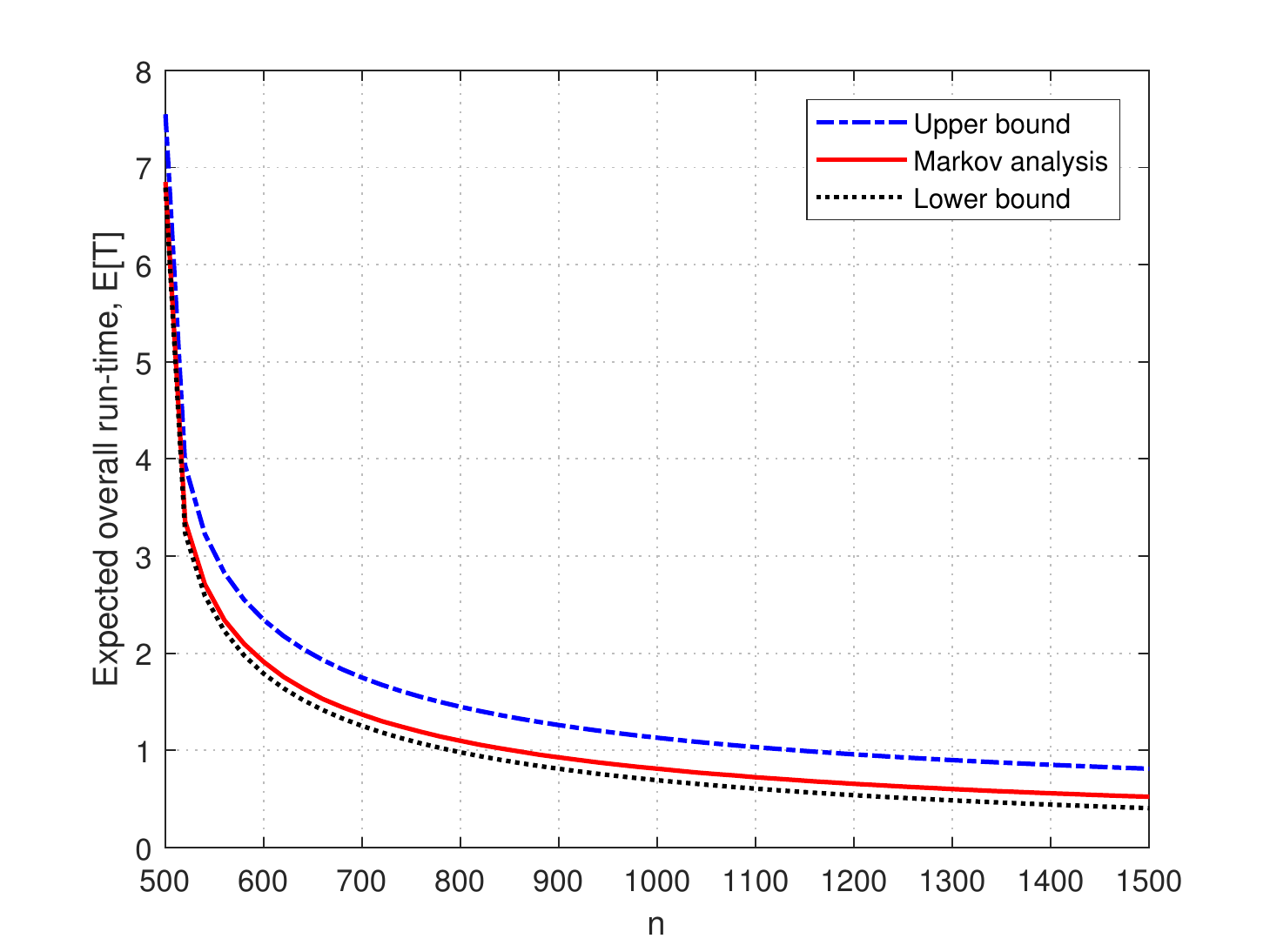}}
	\caption{Bounds of $\mathbb{E}[T]$ for $k=m=500$, $\epsilon=0.1$.}
	\label{fig:latency_versus_n}
\end{figure}

Assume a practical scenario where the size of the task scales linearly with the number of workers, i.e., $k=\Theta(n)$. Note that $k=m$ in this section. Then for large $k$, we have $H_n\simeq \text{log}(n)+\eta$ for a fixed constant $\eta$, which leads to $H_n-H_{n-k}\simeq\text{log}\frac{n}{n-k}$. Thus, the lower bound can be rewritten as 
\begin{equation}\label{eq:Lower_rwrite}
\mathcal{L}=\begin{dcases}
\frac{1}{n\mu_1}+\frac{1}{(1-\epsilon)\mu_2}\text{log}(\frac{n}{n-k}), \ \ \  \epsilon\geq 1 - \frac{\mu_1}{\mu_2} \\
\frac{1}{\mu_1}\text{log}(\frac{n}{n-k})+\frac{1}{n(1-\epsilon)\mu_2}, \ \ \ otherwise,
\end{dcases}
\end{equation}
as $k$ grows. It can be easily shown that $\mathcal{L}=\Theta(1)$. More specifically, assuming a fixed code rate $R=\frac{k}{n}$, the expression of (\ref{eq:Lower_rwrite}) settles down to $\mathcal{L}=\frac{1}{(1-\epsilon)\mu_2}\text{log}(\frac{1}{1-R})$, if $\epsilon\geq 1-\frac{\mu_1}{\mu_2}$ and $\mathcal{L}=\frac{1}{\mu_1}\text{log}(\frac{1}{1-R})$, $otherwise$, as $n$ goes to infinity. 

For the upper bound, we first define 
\begin{equation}\label{eq:fu}
f_U= \frac{H_n-H_{i-k}}{\mu_1}+\frac{H_n-H_{n-i}}{(1-\epsilon)\mu_2}.
\end{equation}
From $\frac{\partial f_{U}}{\partial i}=0$, $f_U$ is minimized by setting $i=i^*$, where 
\begin{equation}
i^*=\frac{\mu_1k+(1-\epsilon)\mu_2n}{\mu_1+(1-\epsilon)\mu_2}.
\end{equation}
Since $i^*-k=\Theta(k)$ and $n-i^*=\Theta(k)$, we can approximate $H_n-H_{i^*-k}\simeq\text{log}\frac{n}{i^*-k}$ and $H_n-H_{n-i^*}\simeq\text{log}\frac{n}{n-i^*}$ for large $k$. Thus, the upper bound in (\ref{eq:Ubound}) can be rewritten as
\begin{equation}
\mathcal{U}= \frac{1}{\mu_1}\text{log}(\frac{n}{i^*-k})+\frac{1}{(1-\epsilon)\mu_2}\text{log}(\frac{n}{n-i^*}).
\end{equation}
It can be again shown that $\mathcal{U}=\Theta(1)$. Finally, we have 
\begin{equation}\label{eq:coded_speed}
\mathbb{E}[T]=\Theta(1). 
\end{equation}
Note that inserting $i=\lceil i^* \rceil$ or $i=\lfloor i^* \rfloor$ at (\ref{eq:fu}) makes the upper bound since $i$ should be an integer. However, the ceiling or floor notation can be ignored for the asymptotic regime. 

\begin{figure}[t]
\centering
     \includegraphics[width=0.41\textwidth, trim=-0.5cm 0 0 0]{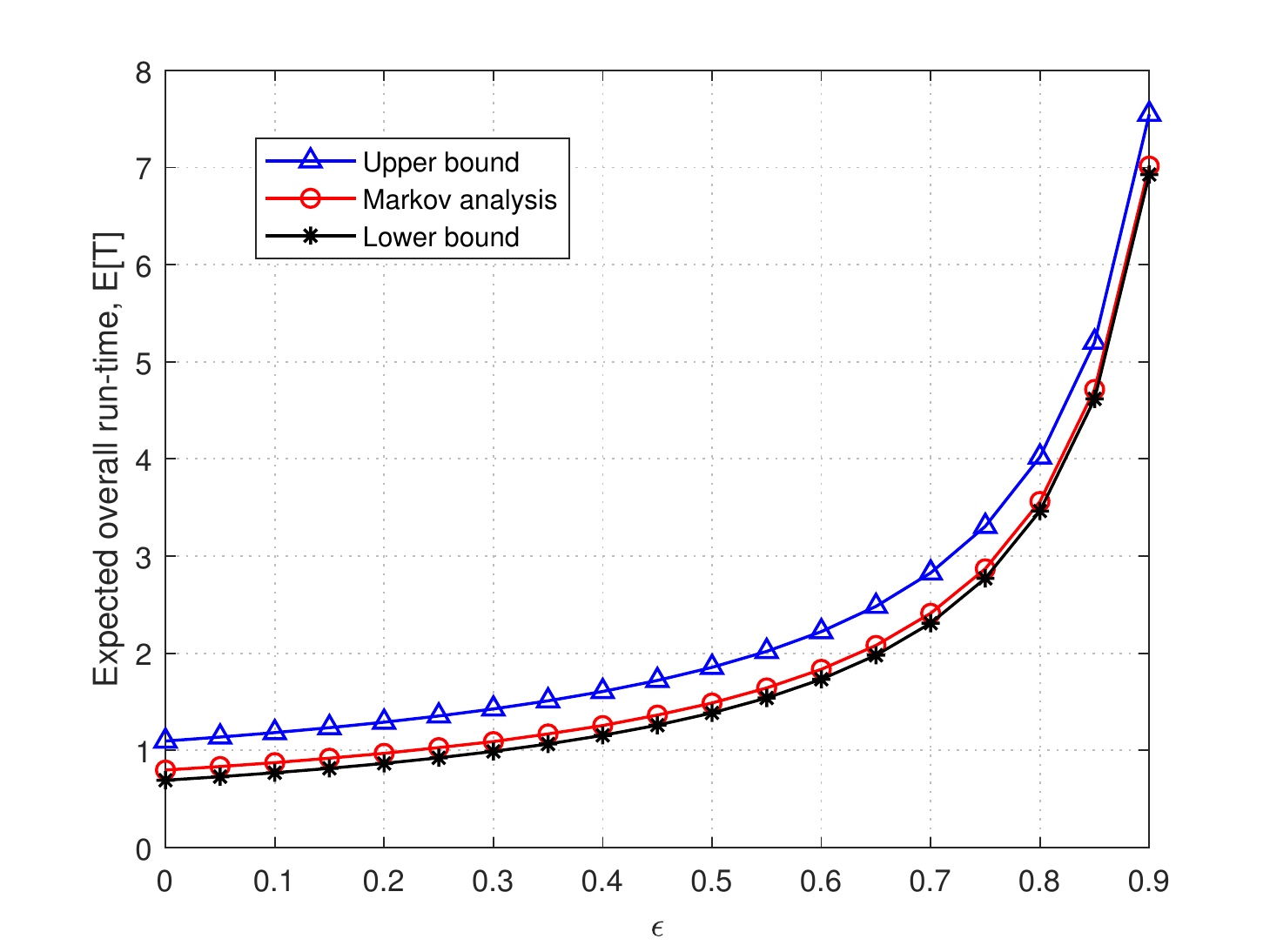}
\caption{Bounds of $\mathbb{E}[T]$ versus $\epsilon$ with $k=m=500$, $\mu_1=10$, $\mu_2=1$.}\label{fig:latancy_versus_epsilon}
\end{figure}


Fig. \ref{fig:latency_versus_n} shows the lower and upper bounds along with the hitting time of the Markov chain, with varying number of workers $n$. The number of inner products to be computed is assumed to be $m=500$. Erasure probability is assumed to be $\epsilon=0.1$. It can be seen that the overall run-time to finish the task decreases as $n$ increases. Based on the Markov analysis and derived bounds, one can find the minimum number of workers $n$ to meet the latency constraint of the given system. 

Fig. \ref{fig:latancy_versus_epsilon} shows the bounds of $\mathbb{E}[T]$ with different erasure probability $\epsilon$. We see by Figs. \ref{fig:latency_versus_n} and \ref{fig:latancy_versus_epsilon} that the lower bound is generally tighter than the upper bound.

\subsection{Comparison with Uncoded Scheme}
In this subsection, we give comparison between the coded and uncoded schemes. For the uncoded scheme, the overall workload ($k=m$ inner products) are equally distributed to $n$ workers. Therefore, the workload at each worker becomes less than a single inner product. More specifically, the number of multiplications and additions computed at each worker are reduced by a factor of $\frac{k}{n}$ compared to the MDS-coded scheme. The master node has to wait for the results from all $n$ workers to finish the task. Since the workload at each worker is decreased by a factor of $\frac{k}{n}$ compared to the coded scheme, the computation time at each worker follows exponential distribution with rate $\frac{n}{k}\mu_1$. For the communication time, we model the communication time for a single inner product at each worker as exponential distribution with rate $\frac{n}{k}\mu_2$, by assuming that the packet length to be transmitted at each worker also decreases by a factor of $\frac{k}{n}$. The erasure probability should be also changed depending on the packet length. Assume that the packet is lost if at least one bit is erroneous. Then, by defining $\epsilon_b$ as the fixed bit-error rate (BER) of the system, the packet erasure probability $\epsilon'$ for the uncoded scheme can be written as  
\begin{equation}
\epsilon'=1-(1-\epsilon_b)^{\frac{k}{n}l}
\end{equation}
where $l$ is the packet length of the coded scheme. Note that the erasure probability for the coded scheme $\epsilon$ is
\begin{equation}
\epsilon=1-(1-\epsilon_b)^{l}.
\end{equation}
The overall run-time of the uncoded scheme denoted by $T_{\text{uncoded}}$ can be written as
\begin{equation}
T_{\text{uncoded}}=\underset{i\in[n]}{\text{max}}(X_i+S_i)
\end{equation}
where $\text{Pr}[X_i\leq t]=1-e^{-\frac{k}{n}\mu_1t}$ and $\text{Pr}[S_i\leq t]=1-e^{-\frac{k}{n}(1-\epsilon)\mu_2t}$ by Lemma 1. Similar to the proof of Theorem 2, we have $\mathcal{L}_{\text{uncoded}}\leq\mathbb{E}[T_{\text{uncoded}}]\leq\mathcal{U}_{\text{uncoded}}$ where
\begin{align*}
\mathcal{L}_{\text{uncoded}}&=\underset{i\in[n]}{\text{max}}(\frac{k(H_n-H_{i-1})}{n\mu_1}+\frac{k(H_n-H_{n-i})}{n(1-\epsilon')\mu_2})\\
&=\begin{dcases}
\frac{k}{n}(\frac{1}{n\mu_1}+\frac{H_n}{(1-\epsilon')\mu_2}), \ \ \ \ \ \ \   \epsilon'\geq 1 - \frac{\mu_1}{\mu_2} \\
\frac{k}{n}(\frac{H_n}{\mu_1}+\frac{1}{n(1-\epsilon')\mu_2}), \ \ \ \ \ \ otherwise,
\end{dcases}
\end{align*}
\begin{equation}
\mathcal{U}_{\text{uncoded}}= \frac{k}{n}(\frac{H_n}{\mu_1}+\frac{H_n}{(1-\epsilon')\mu_2}).
\end{equation}
It can be easily seen that $\mathcal{L}_{\text{uncoded}}=\Theta(\text{log}(n))$, $\mathcal{U}_{\text{uncoded}}=\Theta(\text{log}(n))$ for $k=m=\Theta(n)$, which leads to 
\begin{equation}\label{eq:uncoded_speed}
\mathbb{E}[T_{\text{uncoded}}]=\Theta(\text{log}(n)).
\end{equation}
Comparing (\ref{eq:coded_speed}) and (\ref{eq:uncoded_speed}) shows that the overall run-time becomes $\Theta(\text{log}(n))$ times faster by introducing redundancy:
\begin{equation}
\frac{\mathbb{E}[T_{\text{uncoded}}]}{\mathbb{E}[T]}=\Theta(\text{log}(n)).
\end{equation}

This result is consistent with the previous works \cite{KLee1, heterogeneous} which compared coded and uncoded schemes in an error-free scenario.



\section{Latency for General $k$}\label{sec:analysis2}
In this section, we give discussions for general $k$. We provide the following theorem, which shows the lower and upper bounds of $\mathbb{E}[T]$ for general $k$.
\begin{theorem}
Let $G_i$ be an erlang random variable with shape parameter $\frac{m}{k}$ and rate parameter $1$, i.e., $\text{Pr}[G_i\leq t]=1-\sum_{p=0}^{\frac{m}{k}-1}\frac{1}{p!}e^{-t}t^p 
$. Then for general $k$, we have $\mathcal{L}\leq\mathbb{E}[T]\leq\mathcal{U}$ where
\begin{equation}\label{eq:Lbound_generalk}
\mathcal{L}=\underset{i\in[k]}{\text{max}}\Big(\frac{m(H_n-H_{n-k+i-1})}{k\mu_1}+\frac{\mathbb{E}[G_{(i)}]}{(1-\epsilon)\mu_2}\Big)
\end{equation}
\begin{equation}\label{eq:Ubound_generalk}
\mathcal{U}=\underset{i\in[n]\setminus[k-1]}{\text{min}}\Big(\frac{m(H_n-H_{i-k})}{k\mu_1}+\frac{\mathbb{E}[G_{(i)}]}{(1-\epsilon)\mu_2}\Big).
\end{equation}
\end{theorem}
\begin{IEEEproof}
See Appendix C.
\end{IEEEproof}
\begin{figure*}[!t]
\normalsize
\setcounter{mytempeqncnt}{2}
\begin{equation}\label{eq:erlangmean}
\mathbb{E}[G_{(i)}]=\frac{n!}{(i-1)!(n-i)!\Gamma(\frac{m}{k})}\sum_{p=0}^{i-1}(-1)^p{{i-1}\choose{p}}\sum_{q=0}^{(\frac{m}{k}-1)(n-i+p)}a_q(\frac{m}{k},n-i+p)\frac{\Gamma(\frac{m}{k}+q+1)}{(n-i+p+1)^{\frac{m}{k}+q+1}}
\end{equation}
	\hrulefill
	\vspace*{4pt}
\end{figure*}
As illustrated in \cite{Gamma}, the expected value of $i^{\text{th}}$ order statistic $\mathbb{E}[G_{(i)}]$ can be written as (\ref{eq:erlangmean}) shown at the top of next page. Here, $\Gamma$ is a gamma function\footnote{For a positive integer $z$, $\Gamma(z)=(z-1)!$ is satisfied.} and $a_q(x,y)$ is the coefficient of $t^{q}$ in the expansion of $(\sum_{j=0}^{x-1}\frac{t^{j}}{j!})^{y}$ for any integer $x$, $y$. 

For the uncoded scheme, we have $\mathcal{L}_{\text{uncoded}}\leq\mathbb{E}[T_{\text{uncoded}}]\leq\mathcal{U}_{\text{uncoded}}$ where
 \begin{equation}
\mathcal{L}_{\text{uncoded}}=\underset{i\in[n]}{\text{max}}\Big(\frac{m(H_n-H_{i-1})}{n\mu_1}+\frac{\mathbb{E}[G'_{(i)}]}{(1-\epsilon)\mu_2}\Big)
\end{equation}
\begin{equation}
\mathcal{U}_{\text{uncoded}}=\frac{m(H_n-H_{n-k})}{n\mu_1}+\frac{\mathbb{E}[G'_{(i)}]}{(1-\epsilon)\mu_2}
\end{equation}
which are obtained by inserting $k=n$ at (\ref{eq:Lbound_generalk}) and (\ref{eq:Ubound_generalk}), respectively. Here, $G'_i$ is an erlang random variable with shape parameter $\frac{m}{n}$ and rate parameter $1$, where $\mathbb{E}[G'_{(i)}]$ can be obtained by inserting $k=n$ at (\ref{eq:erlangmean}).
\begin{figure}[t]
\centering
     \includegraphics[width=0.41\textwidth, trim=-0.5cm 0 0 0]{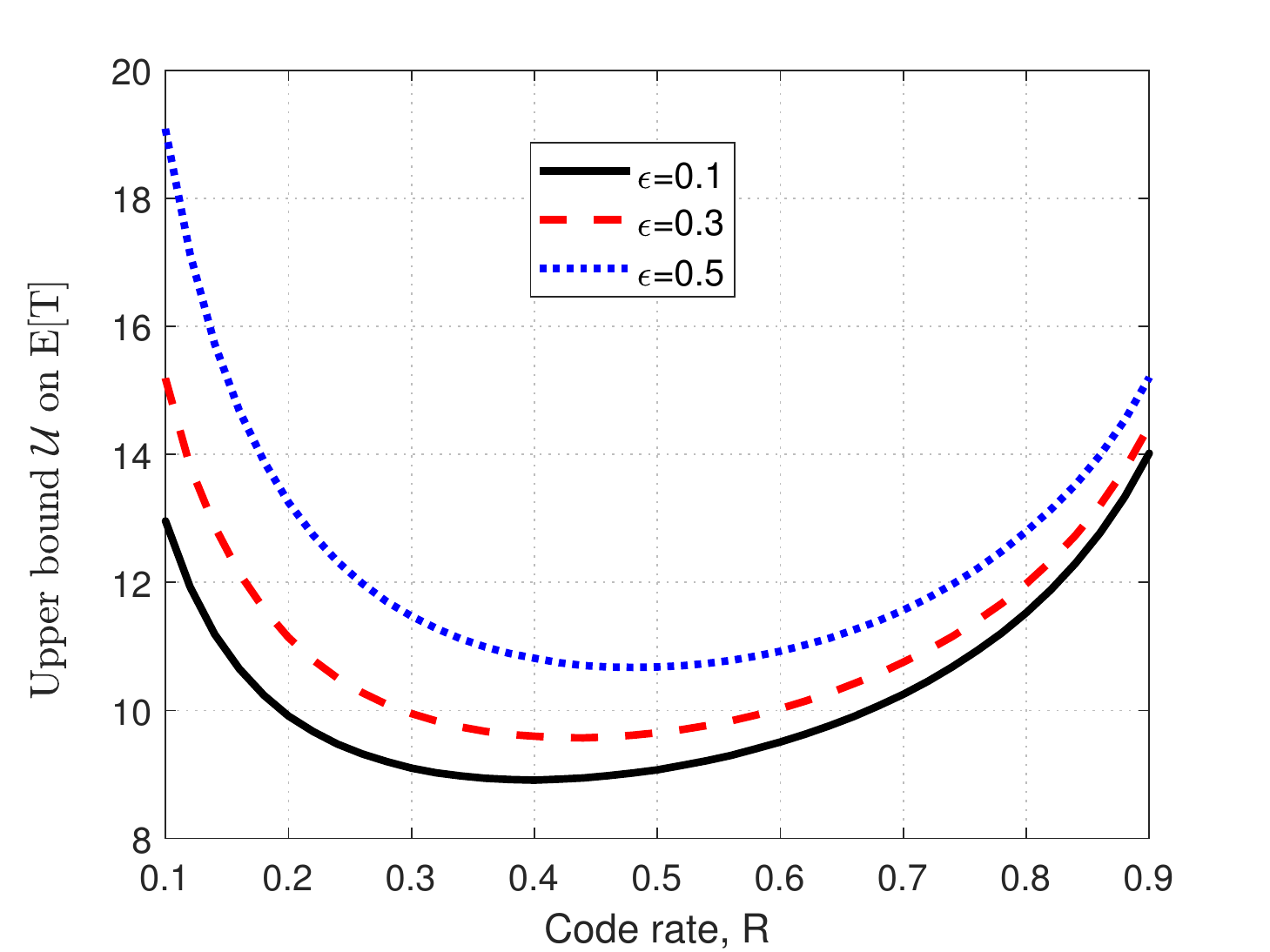}
\caption{Upper bound $\mathcal{U}$ on $\mathbb{E}[T]$ versus code rate $R$ with $n=100$, $m=500$, $\mu_1=1$, $\mu_2=10$.}\label{fig:Optimal_k}
\end{figure}

\begin{figure}[t]
\centering
     \includegraphics[width=0.41\textwidth, trim=-0.5cm 0 0 0]{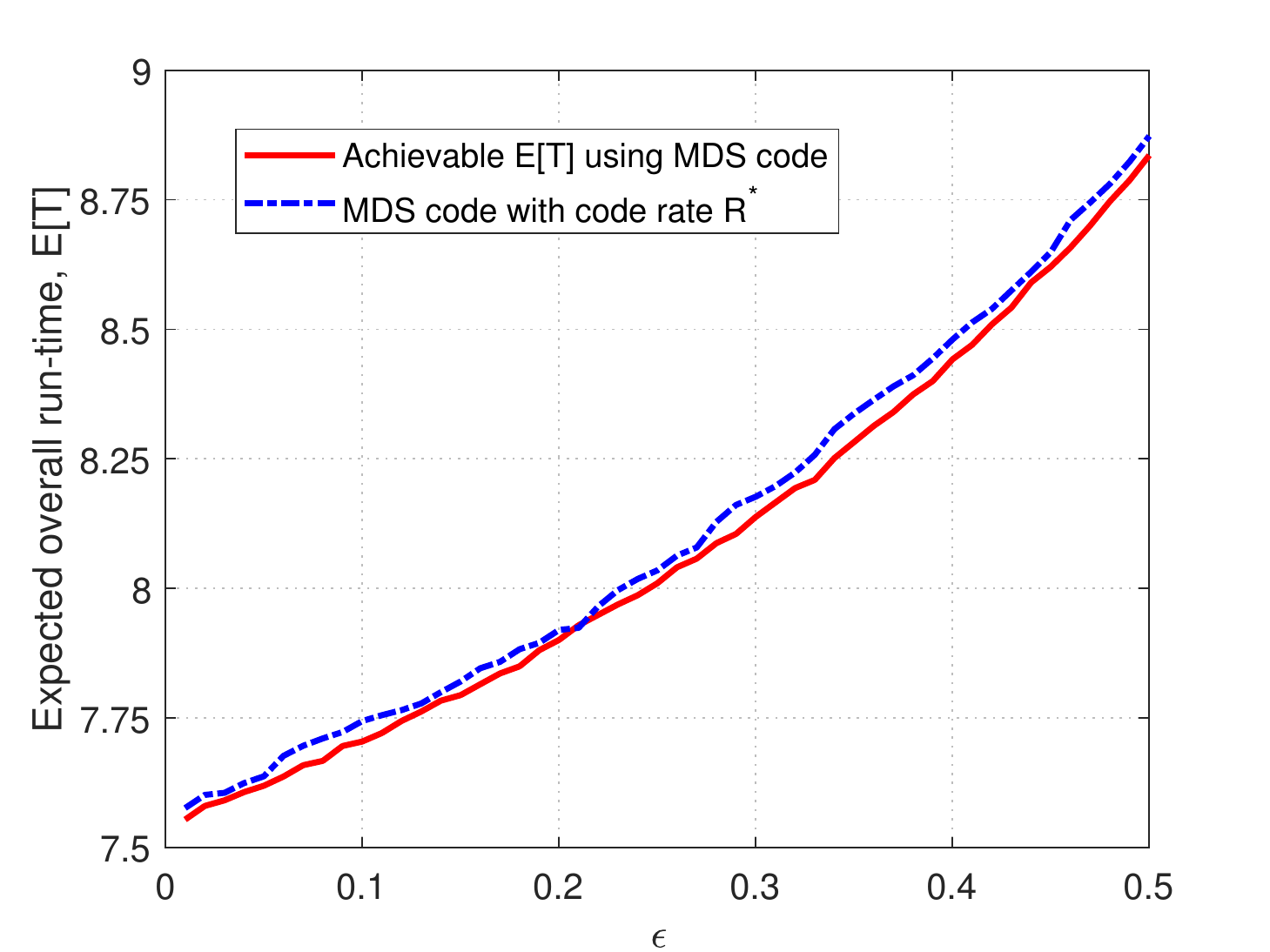}
\caption{Latency performance on the design targeting upper bound minimization. Here, we assume $n=100$, $m=500$, $\mu_1=1$, $\mu_2=10$.}\label{fig:tightness}
\end{figure}


While finding the exact expression of $\mathbb{E}[T]$ for general $k$ remains open, we provide numerical results to gain insights on designing $(n,k)$ MDS code. We plot the upper bound of $\mathbb{E}[T]$ versus code rate $R=\frac{k}{n}$ with $n=100$, $m=500$, $\mu_1=1$, $\mu_2=10$ in Fig. \ref{fig:Optimal_k}. It can be seen that the optimal code rate $R^*$ can be designed to minimize the upper bound $\mathcal{U}$ on $\mathbb{E}[T]$ depending on the erasure probability $\epsilon$. Fig. \ref{fig:tightness} compares the performance between the achievable $\mathbb{E}[T]$ (obtained by Monte Carlo simulations) and the MDS code with rate $R^*$. It can be seen that the performance of the MDS code with rate $R^*$ is very close to the achievable $\mathbb{E}[T]$, showing the effectiveness of the proposed design.

\section{Analysis under the Setting of Limited Number of Retransmissions}\label{sec:Finite}
\subsection{Probability of Successful Computation}
So far, we focused on latency analysis over packet erasure channels under the assumption that there is no limit on the number of transmissions at each worker. The number of transmissions can be potentially infinite on the analysis above. However, in practice, the number of transmissions cannot be potentially infinite due to the constraint on the transmission bandwidth. With limited number of transmissions, the overall computation can fail, i.e., the master node may not receive enough results (less than $k$) from the workers due to link failures.

Thus, we consider the success probability of a given task in this section. We first analyze probability of success at a specific worker. Let $\gamma$ be the maximum number of packets that is allowed to be transmitted at a specific worker. That is, the number of transmissions at each worker is limited to $\gamma$. When the number of workers is given by $n$, the total transmission bandwidth of the network is then limited to $n\gamma l$, where $l$ is the packet length. For a successful computation at a specific worker node, the master node should successfully receive $\frac{m}{k}$ out of $\gamma$ packets from that worker. Thus, probability of success at a specific worker denoted by $p$ is obtained as

\begin{align}
p&=\underbrace{(1-\epsilon)^{\frac{m}{k}}}_{0 \ \text{failure}} + \underbrace{{{\frac{m}{k}}\choose{1}}(1-\epsilon)^{\frac{m}{k}}\epsilon}_{1 \ \text{failure}} + ... \underbrace{{{\gamma-1}\choose{\gamma-\frac{m}{k}}}(1-\epsilon)^{\frac{m}{k}}\epsilon^{\gamma-\frac{m}{k}}}_{\gamma-\frac{m}{k} \ \text{failures}} \nonumber \\
&=\sum_{i=0}^{\gamma-\frac{m}{k}} {{\frac{m}{k}+i-1}\choose{i}}(1-\epsilon)^{\frac{m}{k}}\epsilon^{i}. \label{eq:p}
\end{align} 
To complete the overall task, $k$ out of $n$ workers should successfully deliver their results to the master. Based on $p$, the probability of success for the overall computation, denoted by $P_s$, can be written as 
\begin{equation}\label{eq:P_s}
P_s = \sum_{i=k}^{n}{{n}\choose{i}}(1-p)^{n-i}p^i.
\end{equation}
\begin{figure}[t]
\centering
     \includegraphics[width=0.43\textwidth, trim=-0.5cm 0 0 0]{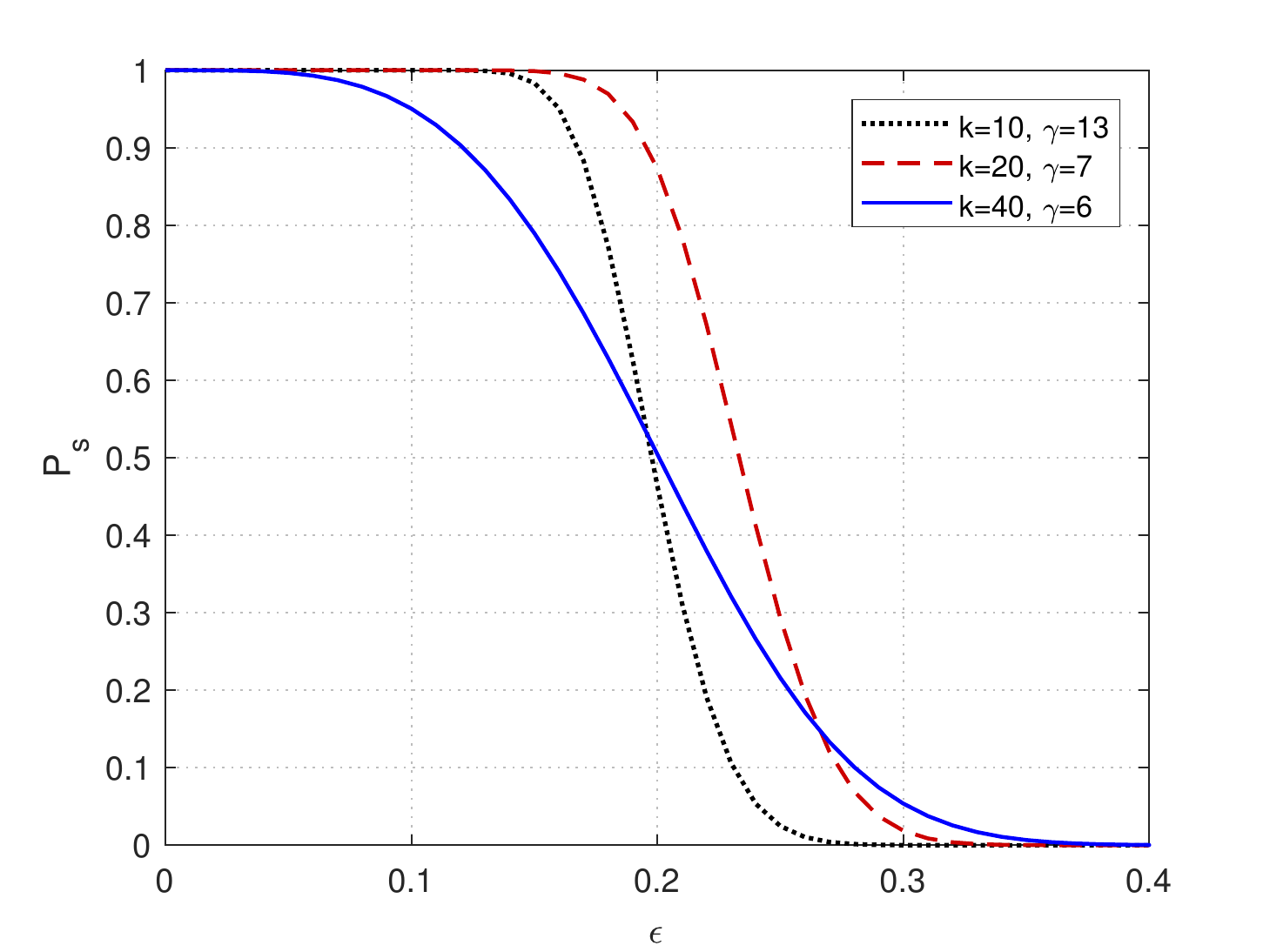}
\caption{$P_s$ versus $\epsilon$ assuming $n=40$, $m=120$. The scheme which uses less transmission bandwidth can have higher probability of success. Moreover, the scheme with higher probability of success in one region may have lower probability of success in the other region.}\label{fig:Ps_vs_epsilon}
\end{figure}
It can be seen from (\ref{eq:p}) that increasing $k$ also increases probability of success $p$ at each worker. This is because increasing $k$ decreases the number of packets to be transmitted at each worker, which increases $p$ for a given number of transmissions $\gamma$. However, larger $k$ increases the number of results to be detected at the master node for decoding the overall computational results. That is, for a given $p$, increasing $k$ decreases $P_s$ from (\ref{eq:P_s}). Now the question is, how to design optimal $k$? How about choosing the appropriate value of $\gamma$? We address these issues in Section V-C. 

For given number of workers $n=40$ and workload $m=120$, Fig. \ref{fig:Ps_vs_epsilon} shows $P_s$ versus erasure probability $\epsilon$ with different $k$ and $\gamma$ values. It can be seen that even the case with $k=20$ uses less bandwidth compared to the case with $k=10$, it has strictly better probability of success for all $\epsilon$ values. This is because the case with $k=10$ has more packets to be transmitted compared to the case with $k=20$. It can be also seen that for $\epsilon>0.2$, the case with $k=40$ has higher $P_s$ compared to the case with $k=10$ which uses twice as many transmission bandwidth as the case with $k=40$. This example motivates us to consider the problem of choosing the best $k$ and $\gamma$ for different $\epsilon$ values under bandwidth and probability of success constraints.

\subsection{Latency}
The overall run-time to finish the task should be reconsidered with limited number of transmissions. Let $T'$ be the overall run-time for a given $\gamma$. Note that if the computation fails for a given $\gamma$, we say that the overall run-time is infinity. Thus, if $P_s<1$, the expected overall run-time $\mathbb{E}[T']$ always becomes infinity. Therefore, in this setup, instead of the expected value $\mathbb{E}[T']$, it is more reasonable to consider probability that the task would be finished within a certain required time $\tau$, $\text{Pr}[T'\leq\tau]$.

We first rewrite $T'$ from (\ref{eq:T}) for a given $\gamma$. Since the term $\gamma$ only gives effect on the communication time at each worker (not the computation time), $T'$ can be written as


\begin{equation}
T'=\underset{i\in[n]}{k^{\text{th}}\text{min}}(X_i+S_i').
\end{equation}
As in (\ref{eq:T}), the computation time at worker $i$ denoted by $X_i$, follows exponential distribution with rate $\mu_1$. The communication time at worker $i$ denoted by $S_i'$ is written as 
\begin{equation}
S_i'=\sum_{j=1}^{C_i}Y_{j},
\end{equation}
where $Y_j$ follows exponential distribution with rate $\mu_2$ and $C_i$ is the number of transmissions at a specific worker until it successfully transmit $\frac{m}{k}$ number of packets, which are all independent for different workers. With probability $1-p$, the worker fails to transmit the $\frac{m}{k}$ packets so that $C_i$ is undefined and $S_i'$ becomes infinity. Thus, we have random variable $C_i$ as follows: 
\begin{equation}
C_i= \begin{dcases}
\frac{m}{k} \ \ \ \ \ \ \ \ \ \text{w.p.} \ (1-\epsilon)^{\frac{m}{k}} \\
\frac{m}{k}+1 \ \ \ \ \text{w.p.} \ {{\frac{m}{k}}\choose{1}}(1-\epsilon)^{\frac{m}{k}}\epsilon}_{ \\
\vdotswithin {\text{erlang}(\frac{m}{k}+1, \mu_2) \ \text{w.p.}}\\
\gamma \ \ \ \ \ \ \ \ \ \ \text{w.p.} \ {{\gamma-1}\choose{\gamma-\frac{m}{k}}}(1-\epsilon)^{\frac{m}{k}}\epsilon^{\gamma-\frac{m}{k}}}_{ \\
\text{undefined} \ \text{w.p.} \ 1-p.
\end{dcases}\end{equation}

Based on $C_i$, random variable $S_i'$ becomes  
\begin{equation}
S_i' \begin{dcases}
\sim\text{erlang}(\frac{m}{k}, \mu_2) \ \text{w.p.} \ (1-\epsilon)^{\frac{m}{k}} \\
\sim\text{erlang}(\frac{m}{k}+1, \mu_2) \ \text{w.p.} \ {{\frac{m}{k}}\choose{1}}(1-\epsilon)^{\frac{m}{k}}\epsilon}_{ \\
\vdotswithin {\text{erlang}(\frac{m}{k}+1, \mu_2) \ \text{w.p.}}\\
\sim\text{erlang}(\gamma, \mu_2) \ \text{w.p.} \ {{\gamma-1}\choose{\gamma-\frac{m}{k}}}(1-\epsilon)^{\frac{m}{k}}\epsilon^{\gamma-\frac{m}{k}}}_{ \\
=\infty \ \text{w.p.} \ 1-p.
\end{dcases}\end{equation}
\begin{figure}[t]
\centering
     \includegraphics[width=0.43\textwidth, trim=-0.5cm 0 0 0]{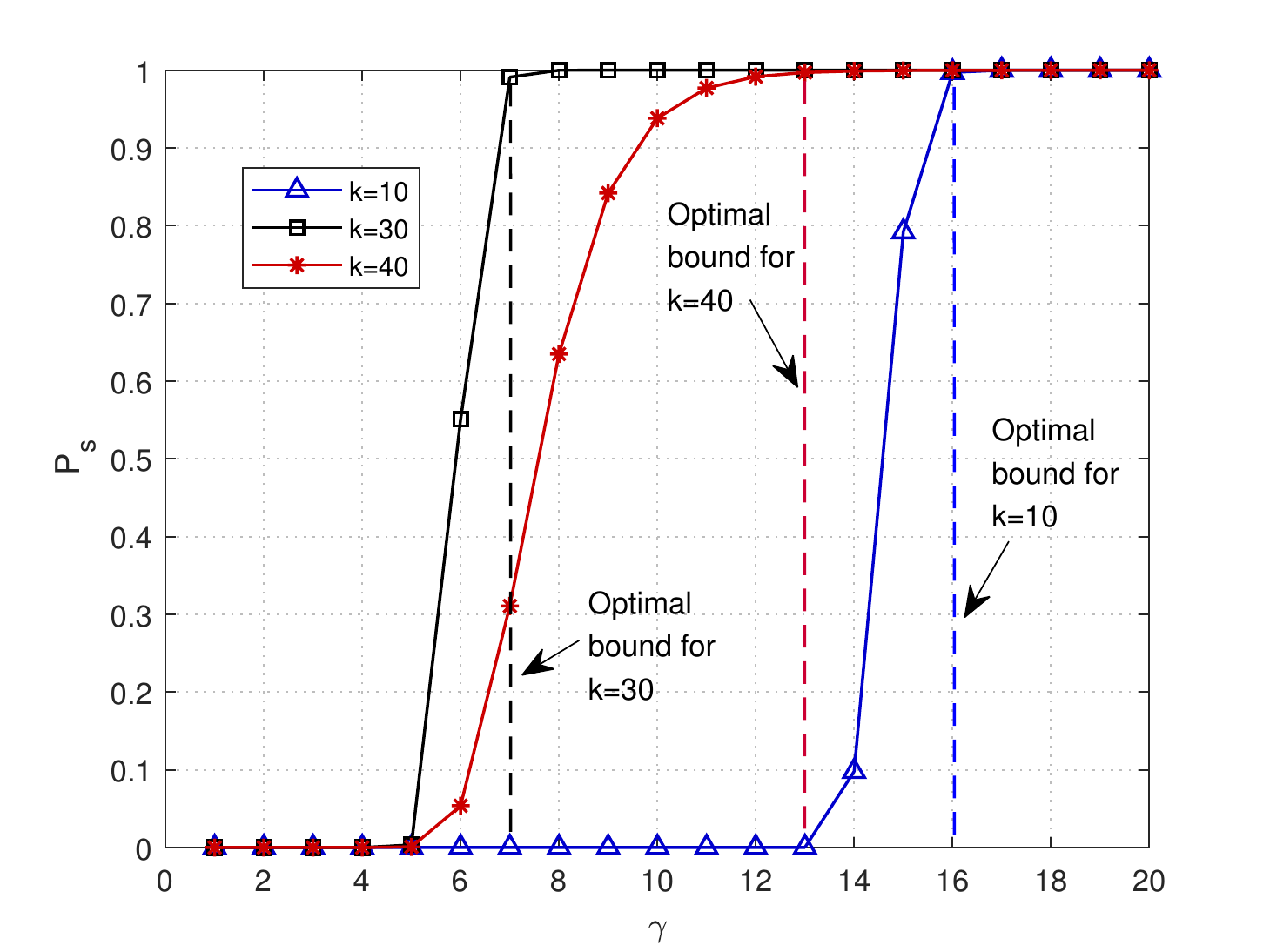}
\caption{$P_s$ versus $\gamma$ assuming $n=40$, $m=120$, $\epsilon=0.3$. The figure shows the minimum $\gamma$ satisfying $P_s\geq 1-\delta$ for different $k$ values, where $\delta$ is set to $0.01$.}\label{fig:Ps_vs_gamma}
\end{figure}
\textit{Remark} 2: As $\gamma$ increases, the probability $\text{Pr}[T'\leq \tau]$ also increases for a given $\tau$, since increasing $\gamma$ increases the probability of success. The probability $\text{Pr}[T'\leq \tau]$ converges to $\text{Pr}[T\leq \tau]$ as $\gamma$ increases, where $T$ is the asymptotic overall run-time defined in (\ref{eq:T}).

\subsection{Optimal Resource Allocation in Practical Scenarios}

For given constraints on maximum number of transmissions $\gamma$ and probability of successful computation $P_s$, we would like to minimize $T^{(\alpha)}$ which is defined as the minimum overall run-time that we can guarantee with probability $1-\alpha$, as follows:

\begin{equation}
T^{(\alpha)}=\text{min}\{\tau: \text{Pr}[T'\leq \tau]\geq 1-\alpha \}.
\end{equation} 

\begin{figure}[t]
\centering
     \includegraphics[width=0.43\textwidth, trim=-0.5cm 0 0 0]{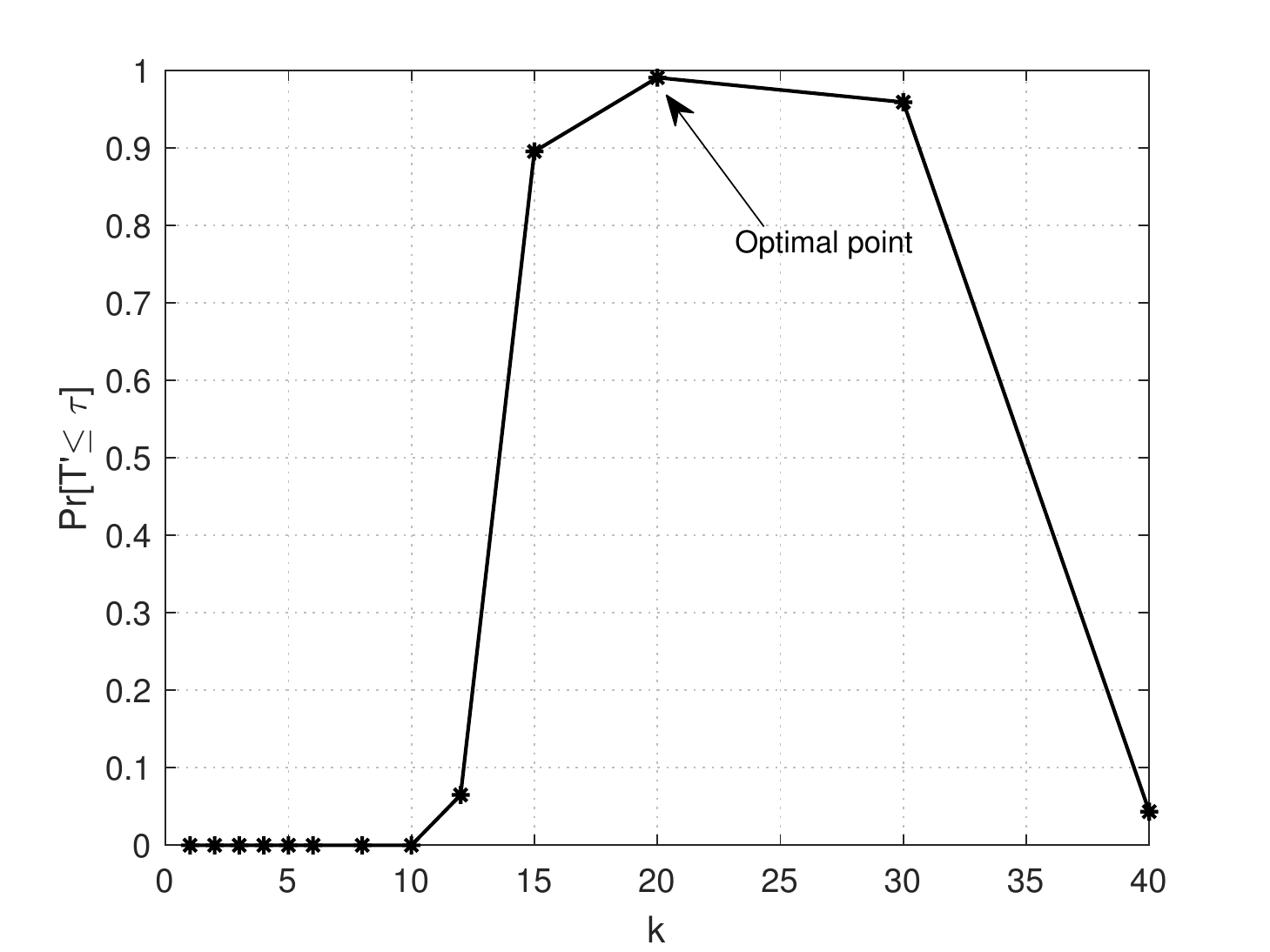}
\caption{$\text{Pr}[T'\leq\tau]$ versus $k$ assuming $n=40$, $m=120$, $\gamma=13$, $\tau=8.6$, $\epsilon=0.3$, $\mu_1=1$, $\mu_2=5$.}\label{fig:Prlatency_vs_k}
\end{figure}

For a given number of workers $n$, the optimization problem is formulated as 
\begin{align}
\underset{k,\gamma}{\text{min}} & \ T^{(\alpha)}  \\
\text{subject  to } & \gamma\leq\gamma_t, \\
& P_s\geq1-\delta,
\end{align}
where $\gamma_t$ is the transmission bandwidth limit and $\delta$ shows how $P_s$ is close to 1 ($0\leq \delta \leq 1$). While it is not straightforward to find the closed-form solution of the above problem, optimal $k$ can be found by combining the results of Figs \ref{fig:Ps_vs_gamma} and \ref{fig:Prlatency_vs_k}. We first observe Fig. \ref{fig:Ps_vs_gamma}, which shows the probability successful computation $P_s$ as a function of $\gamma$. The parameters are set as $n=40$, $m=120$, $\epsilon=0.3$. Obviosly, $P_s$ increases as $\gamma$ increases. In addition, since there are $\frac{m}{k}$ packets to be transmitted at each worker, we have $P_s=0$ for $\gamma<\frac{m}{k}$. The optimal bounds shown in Fig. \ref{fig:Ps_vs_gamma} are the minimum $\gamma$ values that satisfy $P_s\geq 1-\delta$, where $\delta=0.01$. Given a success probability constraint $P_s\geq 1-\delta$ and bandwidth constraint $\gamma\leq\gamma_t$, one can find the candidates of $k$ values that satisfy the constraints. For example, if $\gamma_t=13$, $k=30$ and $k=40$ become possible candidates for the optimal solution while $k=10$ does not. Note that in Fig. \ref{fig:Ps_vs_gamma}, increasing (or decreasing) $k$ does not always indicate higher probability of success.

\begin{figure}[t]
\centering
     \includegraphics[width=0.43\textwidth, trim=-0.5cm 0 0 0]{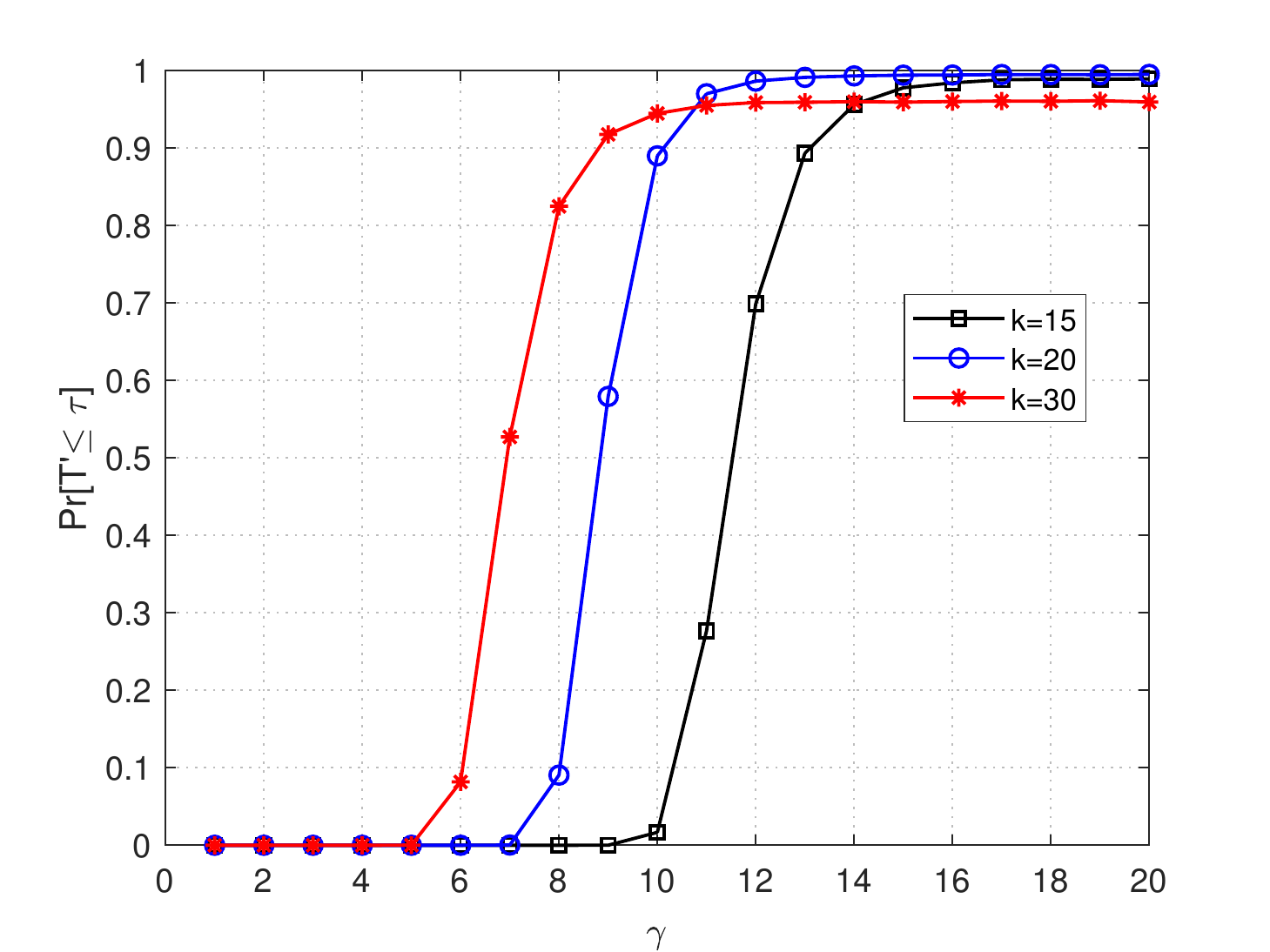}
\caption{$\text{Pr}[T'\leq \tau_t]$ versus $k$ assuming $n=40$, $m=120$, $\tau_t=8.6$, $\epsilon=0.3$, $\mu_1=1$, $\mu_2=5$.}\label{fig:Prlatency_vs_gamma}
\end{figure}

\begin{figure}[t]
\centering
     \includegraphics[width=0.43\textwidth, trim=-0.5cm 0 0 0]{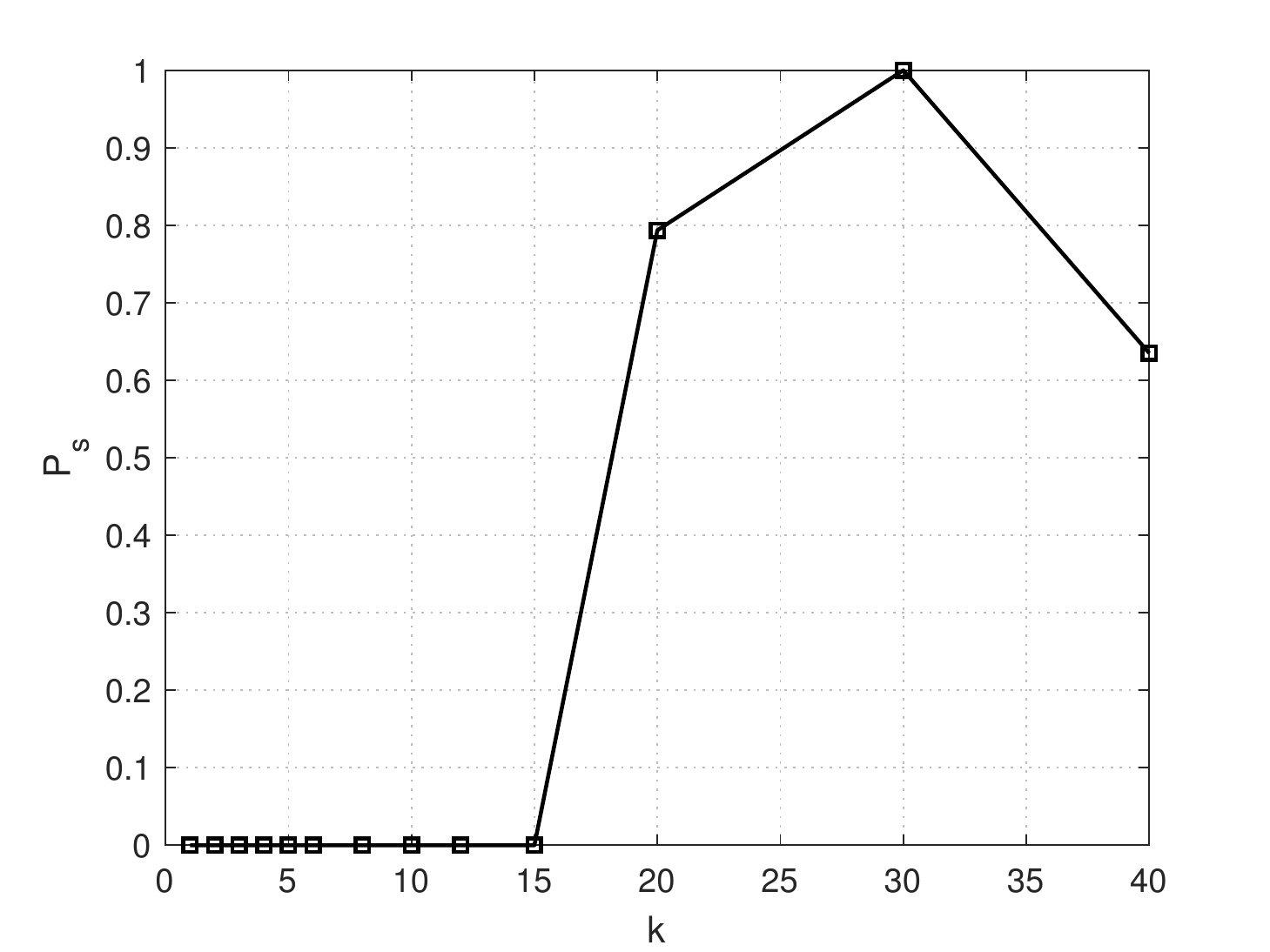}
\caption{$P_s$ versus $k$ assuming $n=40$, $m=120$, $\epsilon=0.3$, $\gamma=8$.}\label{fig:Ps_vs_k}
\end{figure}
\begin{figure}
	\centering
	\subfloat[][Optimal $T^{(\alpha)}-\epsilon$ curve.]{\includegraphics[height=53mm]{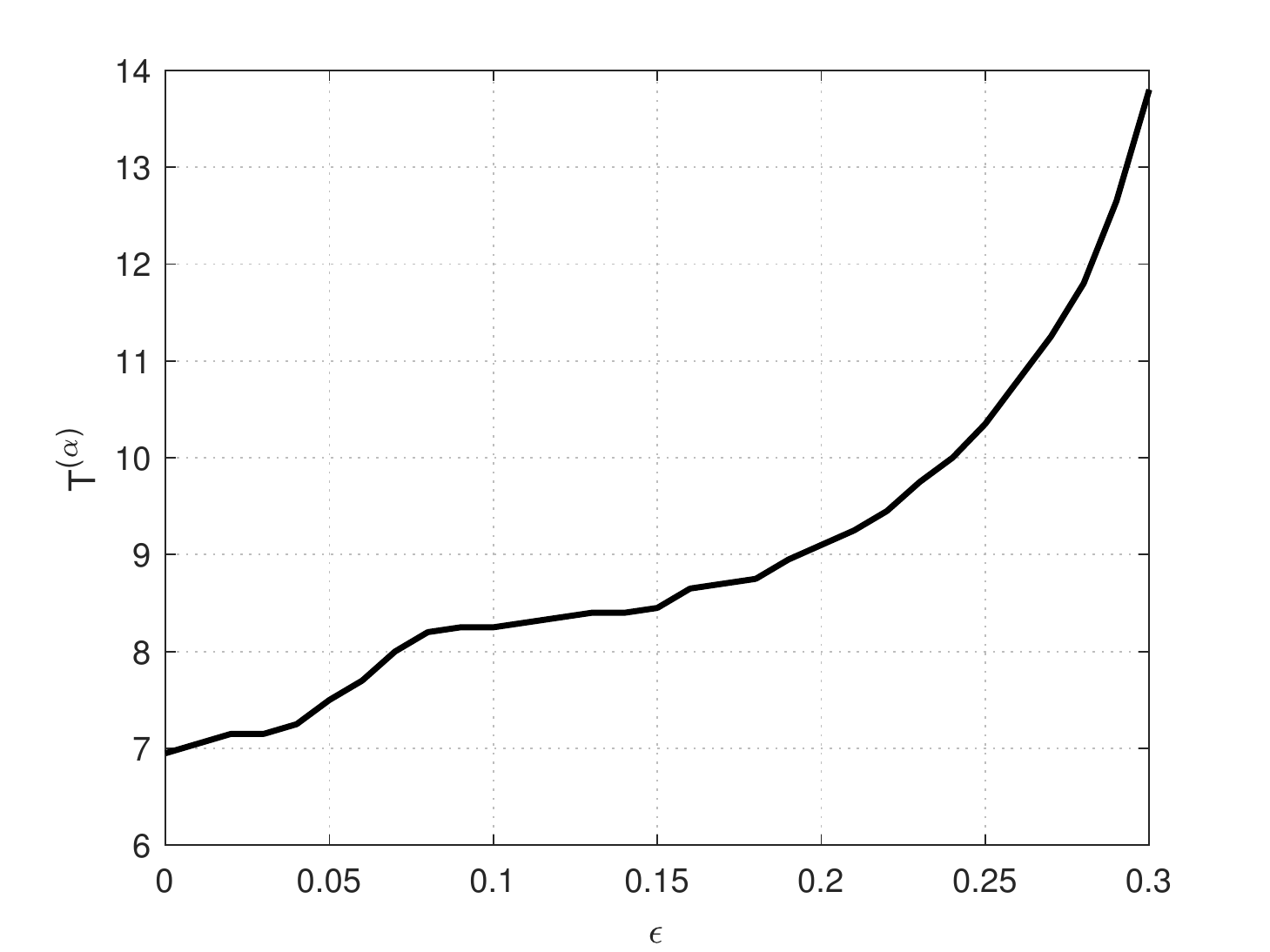}}
	\quad \quad
	\subfloat[][Optimal $\gamma-\epsilon$ curve.]{\includegraphics[height=53mm]{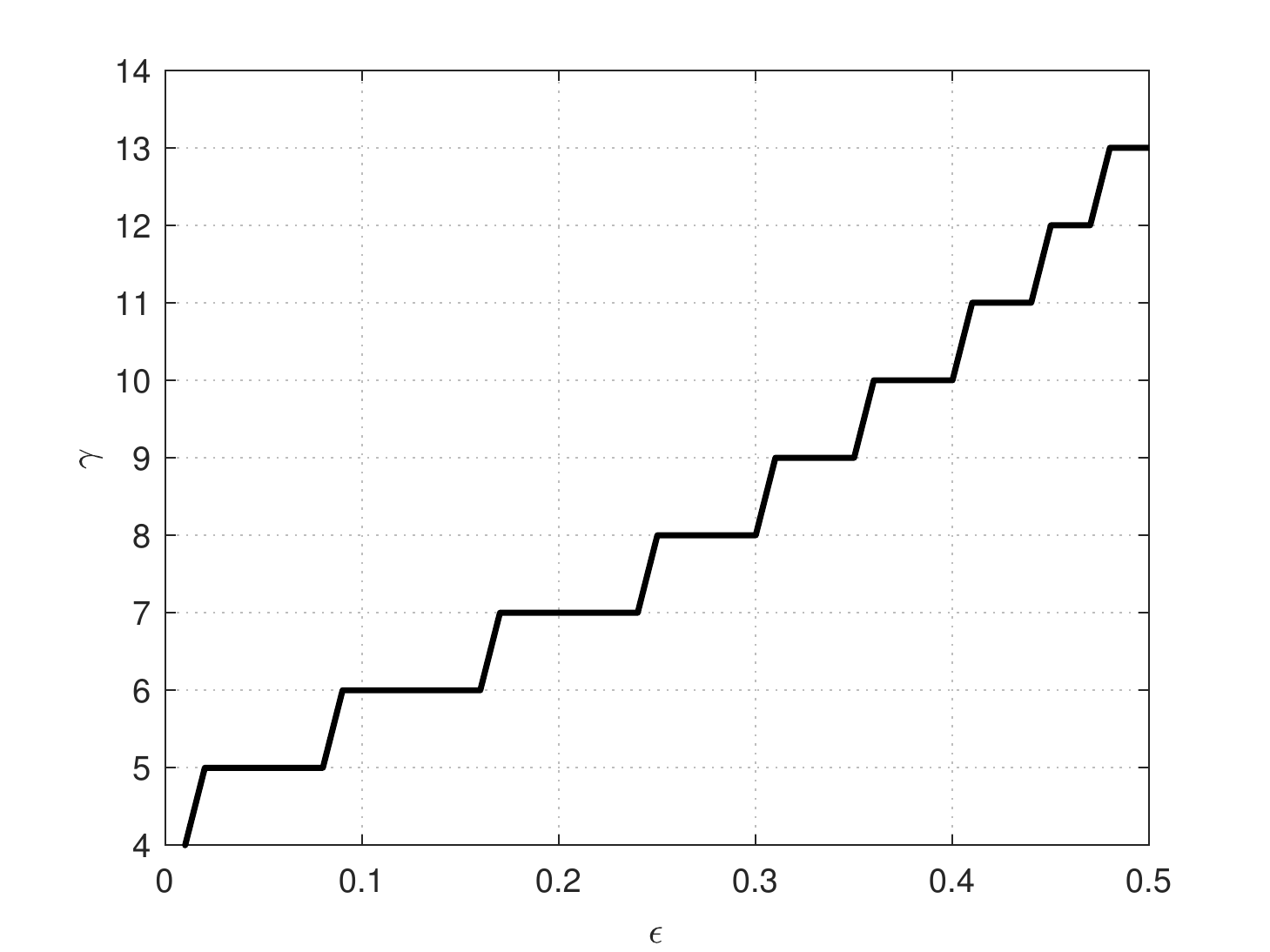}}
   \quad \quad
	\subfloat[][Optimal $P_s-\epsilon$ curve.]{\includegraphics[height=53mm]{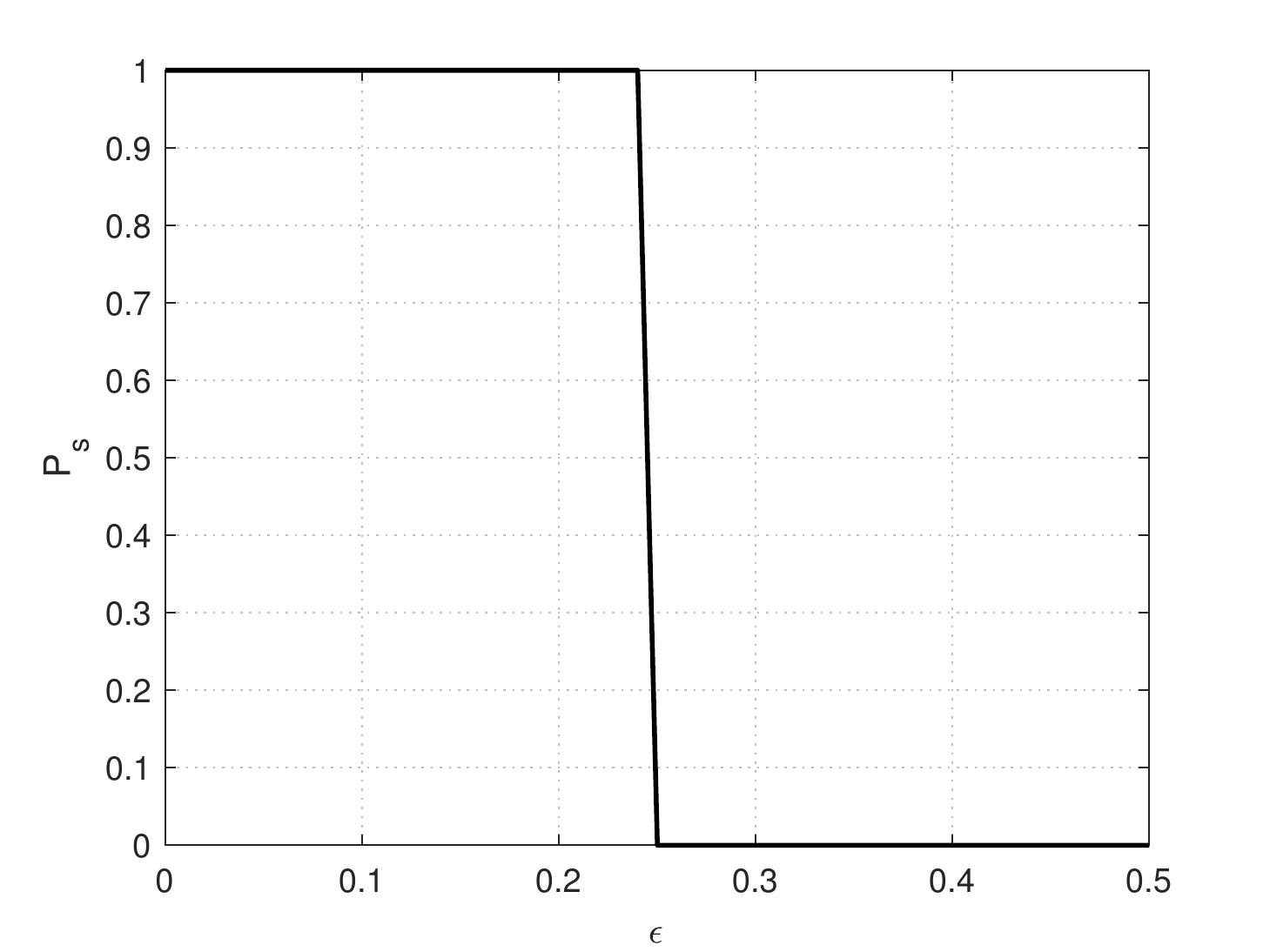}}
	\caption{Optimal curves assuming $n=40$, $m=120$, $\mu_1=1$, $\mu_2=5$, $\gamma_t=7$, $\tau_t=10$, $\alpha=0.05$, $\delta=0.01$.}
	\label{fig:Optimal_curve}
\end{figure}
Fig. \ref{fig:Prlatency_vs_k} shows $\text{Pr}[T'\leq \tau]$ versus $k$ assuming $\gamma=13$, $\tau=8.6$, where other parameters are set to be same as in Fig. \ref{fig:Ps_vs_gamma}. Among the candidates obtained from Fig. \ref{fig:Ps_vs_gamma}, one can find the optimal $k$ that minimizes $T^{(\alpha)}$. By setting $\alpha=0.03$ and $\tau=8.6$, it can be seen from Fig. \ref{fig:Prlatency_vs_k} that $\text{Pr}[T'\leq \tau]\geq 1-\alpha$ holds only for $k=20$, which means that $k=20$ is the optimal solution. Note that $k$ that minimizes $\gamma$ to satisfy $P_s\geq 1-\delta$ (in Fig. \ref{fig:Ps_vs_gamma}) does not always minimizes $T^{(\alpha)}$, which motivated us to formulate the problem (while it is not shown in Fig. \ref{fig:Ps_vs_gamma}, the case for $k=20$ has lower performance compared to the case for $k=30$).

For given constraints on probability of success and latency, one can also minimize $\gamma$:
\begin{align}
\underset{k}{\text{min}} & \ \gamma  \\
\text{subject  to } & P_s\geq1-\delta, \\
& T^{(\alpha)}\leq \tau_t.
\end{align} 
where $\tau_t$ is the required latency constraint. For this optimization problem, it can be seen that the constraint $T^{(\alpha)}\leq \tau_t$ is equivalent to the constraint $\text{Pr}[T'\leq \tau_t]\geq 1-\alpha$. From Fig. \ref{fig:Ps_vs_gamma}, candidates of $(k,\gamma)$ pairs satisfying $P_s\geq 1-\delta$ can be obtained. Among these candidates, one can find the best $k$ that minimizes $\gamma$ from Fig. \ref{fig:Prlatency_vs_gamma}, which shows $\text{Pr}[T'\leq \tau_t]$ versus $\gamma$ with different $k$ values, where $\tau_t$ is set to $8.6$. With small number of retransmissions, $k=30$ gives the best performance while for $\gamma \geq 10$, it does not. This is because the case with $k=30$ has smaller number of packets to be transmitted compared to others, which gives higher $P_s$ for small $\gamma$. As $\gamma$ grows, the probability $\text{Pr}[T'\leq \tau_t]$ converges to $\text{Pr}[T\leq \tau_t]$ for each $k$.

At last, we formulate the following problem,
\begin{align}
\underset{k,\gamma}{\text{max}} & \ P_s  \\
\text{subject  to } & \gamma\leq\gamma_t, \label{eq:gamma_constraint3}\\
& T^{(\alpha)}\leq \tau_t. \label{eq:latency_constraint3}
\end{align} 
aiming to maximize the probability of success with transmission bandwidth and latency constraints. We first find the candidates of $k$ that satisfy $T^{(\alpha)}\leq\tau_t$ from Fig. \ref{fig:Prlatency_vs_gamma}, and then choose optimal $k$ from Fig. \ref{fig:Ps_vs_k} which shows $P_s$ as a function of $k$.  

\subsection{Achievable Curves}
Based on the solutions to the optimization problems, optimal curves as a function of $\epsilon$ are obtained in Fig. \ref{fig:Optimal_curve}, where the parameters are set as $n=40$, $m=120$, $\mu_1=1$, $\mu_2=5$, $\gamma_t=7$, $\tau_t=10$, $\alpha=0.05$, $\delta=0.01$. Fig. \ref{fig:Optimal_curve}(a) shows the achievable $T^{(\alpha)}$ as a function of $\epsilon$. We observe that the achievable $T^{(\alpha)}$ increases as $\epsilon$ grows. From Fig. \ref{fig:Optimal_curve}(b) which shows the optimal $\gamma-\epsilon$ curve, we observe that achievable $\gamma$ also increases as $\epsilon$ grows, which indicates that a larger transmission bandwidth is required for higher erasure probability. Finally in Fig. \ref{fig:Optimal_curve}(c), we can see that there exists a sharp threshold $\epsilon_t$ that makes $P_s=0$ for $\epsilon>\epsilon_t$ since the latency constraint cannot be satisfied anymore for $\epsilon$ greater than some threshold $\epsilon_t$.  

\section{Concluding Remarks}\label{sec:conclusion}
We studied the problem of coded computation assuming link failures, which are a fact of life in current wired data centers and wireless networks. With maximum $k$, we showed that the asymptotic expected latency is equivalent to analyzing a continuous-time Markov chain. Closed-form expressions of lower and upper bounds on the expected latency are derived based on some key inequalities. For general $k$, we show that the performance of the coding scheme which minimizes the upper bound nearly achieves the optimal run-time, which shows the effectiveness of the proposed design. Finally, under the setting of limited number of retransmissions, we formulated practical optimization problems and obtained achievable curves as a function of packet erasure probability $\epsilon$. Considering the link failures in hierarchical computing systems \cite{HPark} or in secure distributed computing \cite{Secure} are interesting and important topics for future research.







\appendices
\section{Proof of Lemma 1}
Assuming $k=m$ (i.e., $r=1$), we have $S_i=\sum_{j=1}^{N_1}Y_{j,1}$. For fixed integer $q$, $\sum_{j=1}^{q}Y_{j,1}$ which is the sum of $q$ exponential random variables with rate $\mu_2$, follows erlang distribution with shape parameter $q$ and rate parameter $\mu_2$: 
\begin{equation}
\text{Pr}[\sum_{j=1}^{q}Y_{j,1}\leq t]=1-\sum_{p=0}^{q-1}\frac{1}{p!}e^{-\mu_2t}(\mu_2t)^p. 
\end{equation}
Now we can write
\begin{align*}
\text{Pr}[S_i\leq t]&=\sum_{q=1}^{\infty}\text{Pr}[N_1=q]\text{Pr}[\sum_{j=1}^{q}Y_{j,1}\leq t]\\
&=\sum_{q=1}^{\infty}((1-\epsilon)\epsilon^{q-1})(1-\sum_{p=0}^{q-1}\frac{1}{p!}e^{-\mu_2t}(\mu_2t)^p)\\
&=1-e^{-\mu_2t}\sum_{q=1}^{\infty}\sum_{p=0}^{q-1}((1-\epsilon)\epsilon^{q-1})\frac{1}{p!}(\mu_2t)^p\\
&=1-e^{-\mu_2t}\sum_{p=0}^{\infty}\sum_{q=p+1}^{\infty}((1-\epsilon)\epsilon^{q-1})\frac{1}{p!}(\mu_2t)^p\\
&=1-e^{-\mu_2t}\sum_{p=0}^{\infty}\frac{1}{p!}(\epsilon\mu_2t)^p\\
&=1-e^{-(1-\epsilon)\mu_2t}
\end{align*}
which completes the proof.

\section{Proof of Theorem 2}
\subsection{Lower bound}
\begin{figure}[t]
\centering
     \includegraphics[width=0.46\textwidth, trim=-0.5cm 0 0 0]{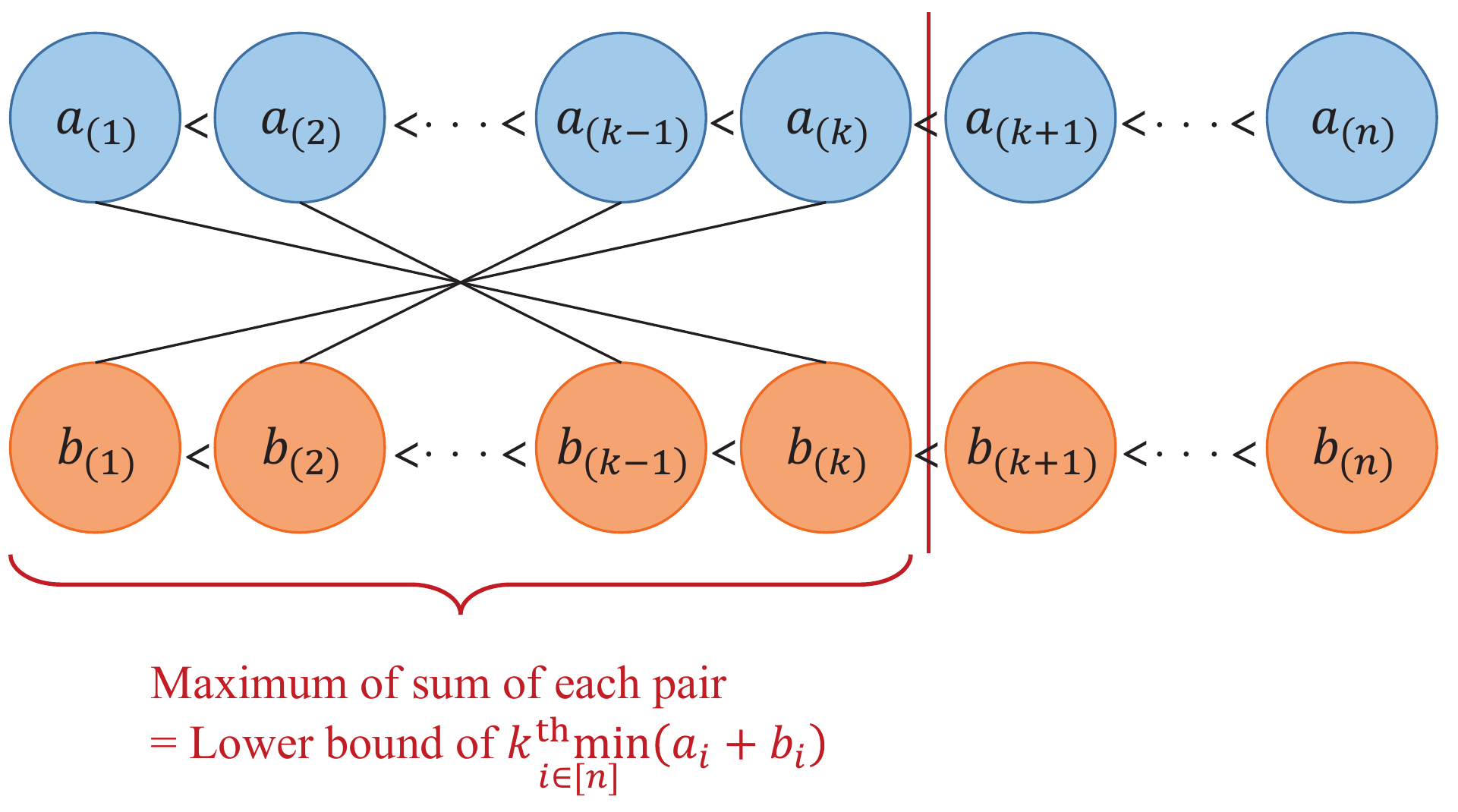}
\caption{Illustration of Proposition 1.}\label{fig:Proposition1}
\end{figure}

We first prove the following proposition, which is illustrated as in Fig. \ref{fig:Proposition1}.
\begin{proposition}
For any two sequences $\{a_i\}_{i\in[n]}$, $\{b_i\}_{i\in[n]}$,
\begin{equation}\label{eq:prop1}
\underset{i\in[n]}{k^{\text{th}}\text{min}}(a_i+b_i)\geq\underset{i\in[k]}{\text{max}}(a_{(k-i+1)}+b_{(i)}).
\end{equation}
\end{proposition}
\begin{IEEEproof}
Define the ordering vector $\boldsymbol{\pi}\in\Pi$, where $\Pi$ is the set of permutation of $[n]$, i.e.,
\begin{equation}\label{eq:ordervector}
\Pi=\{\boldsymbol{\pi}=[\pi_1, ...,\pi_n]: \pi_i\in[n] \ \forall i\in[n], \ \pi_i\neq\pi_j \ for \ i\neq j \}.
\end{equation}
Note that the $k^{\text{th}}$ smallest value out of $n$ can be always rewritten as the maximum of $k$ smallest values. Define $\boldsymbol{\pi}=[\pi_{1}, ...,\pi_{n}]\in\Pi$ to make the set of values $\{a_{\pi_i}+b_{\pi_i}\}_{i=1}^{k}$ be the $k$ smallest among $\{a_i+b_i\}_{i=1}^{n}$. In addition, define $\boldsymbol{\pi}_1,\boldsymbol{\pi}_2\in\Pi$ to make $a_{\pi_i}=a_{(\pi_{i,1})}$ and $b_{\pi_i}=b_{(\pi_{i,1})}$ be satisfied $\forall i\in[k]$, where $\boldsymbol{\pi}_1=[\pi_{1,1}, ...,\pi_{n,1}]$, $\boldsymbol{\pi}_2=[\pi_{1,2}, ...,\pi_{n,2}]$. Then, $\underset{i\in[n]}{k^{\text{th}}\text{min}}(a_i+b_i)$ can be rewritten as   
\begin{align}
\underset{i\in[n]}{k^{\text{th}}\text{min}}(a_i+b_i)&=\underset{i\in[k]}{\text{max}}(a_{\pi_i}+b_{\pi_i}) \label{eq:kthmin_L1}\\
&=\underset{i\in[k]}{\text{max}}(a_{(\pi_{i,1})}+b_{(\pi_{i,2})}). \label{eq:kthmin_L2}
\end{align}
Now define a new set of ordering vectors $\Pi^*\subseteq\Pi$, as $\Pi^*=\{\boldsymbol{\pi}\in\Pi:\pi_i\in[k] \ \forall i\in[k]\}$. Then we can always construct $\boldsymbol{\pi}_1^*, \boldsymbol{\pi}_2^*\in \Pi^*$ such that $\pi_{i,1}\geq\pi_{i,1}^*$, $\pi_{i,2}\geq\pi_{i,2}^*$ for all $i\in [k]$, which is easy to proof. Then by adequately selecting $\boldsymbol{\pi}^*\in \Pi^*$, (\ref{eq:kthmin_L2}) is lower bounded as 
\begin{align}
\underset{i\in[k]}{\text{max}}(a_{(\pi_{i,1})}+b_{(\pi_{i,2})})&\geq\underset{i\in[k]}{\text{max}}(a_{(\pi_{i,1}^*)}+b_{(\pi_{i,2}^*)})\label{eq:L_inequal} \\
&=\underset{i\in[k]}{\text{max}}(a_{(\pi_i^*)}+b_{(i)})\label{eq:L_11}.
\end{align}
Equation (\ref{eq:L_inequal}) implies that the lower bound of $\underset{i\in[k]}{\text{max}}(a_{(\pi_{i,1})}+b_{(\pi_{i,2})})$ can be obtained by properly selecting the $k$ pairs $\{a_{(\pi_{i,1}^*)}+b_{(\pi_{i,2}^*)}\}_{i=1}^{k}$ all in the left region of Fig. \ref{fig:Proposition1}, and taking the maximum. This value $\underset{i\in[k]}{\text{max}}(a_{(\pi_{i,1}^*)}+b_{(\pi_{i,2}^*)})$ can be simply rewritten as (\ref{eq:L_11}). Now we want to show that for any $\boldsymbol{\pi}^*\in\Pi^*$, (\ref{eq:L_11}) cannot be smaller than $\underset{i\in[k]}{\text{max}}(a_{(k-i+1)}+b_{(i)})$, which is the lower bound illustrated in Fig. \ref{fig:Proposition1}.



Define $L$ as
\begin{equation}
L=\underset{i\in[k]}{\text{max}}(a_{(k-i+1)}+b_{(i)})=a_{(k-j+1)}+b_{(j)}
\end{equation}
for some $j\in[k]$. Suppose there exist $L'$ and an ordering vector $\boldsymbol{\pi}'=[\pi_1', ...,\pi_n']\in\Pi^*$ such that
\begin{equation}\label{eq:contr1}
\underset{i\in[k]}{\text{max}}(a_{(\pi_i')}+b_{(i)})=L'<L.
\end{equation}
We will show that the assumption of (\ref{eq:contr1}) is a contradiction, which completes the proof. By (\ref{eq:contr1}), $a_{(\pi_{j'}')}+b_{(j')}<a_{(k-j+1)}+b_{(j)}$ should be satisfied for all $j'\in[k]$. This implies $\pi_{j'}'<k-j+1$ for all $j'\in [k]\setminus[j-1]$, which is a contradiction since one-to-one mapping between $j'$ and $\pi_{j'}'$ cannot be constructed for $j'\in[k]$. 





\end{IEEEproof}

Since $\text{Pr}[X_i\leq t]=1-e^{-\mu_1t}$ and $\text{Pr}[S_i\leq t]=1-e^{-(1-\epsilon)\mu_2t}$ for $k=m$, we have 
\begin{equation}\label{eq:expected_xi}
\mathbb{E}[X_{(i)}]=\frac{H_n-H_{n-i}}{\mu_1} 
\end{equation}
\begin{equation}\label{eq:expected_si}
\mathbb{E}[S_{(i)}]=\frac{H_n-H_{n-i}}{(1-\epsilon)\mu_2}.
\end{equation}
for all $i\in[n]$. Then, the lower bound is derived as 
\begin{align*}
\mathbb{E}[T]&=\mathbb{E}[\underset{i\in[n]}{k^{\text{th}}\text{min}}(X_i+S_i)]\\
&\underset{(a)}{\geq} \mathbb{E}[\underset{i\in[k]}{\text{max}}(X_{(k-i+1)}+S_{(i)})]\\
&\underset{(b)}{\geq} \underset{i\in[k]}{\text{max}}(\frac{H_n-H_{n-k+i-1}}{\mu_1}+\frac{H_n-H_{n-i}}{(1-\epsilon)\mu_2})\\
&\underset{(c)}{=}\begin{dcases}
\frac{H_n-H_{n-1}}{\mu_1}+\frac{H_n-H_{n-k}}{(1-\epsilon)\mu_2}, \ \ \epsilon\geq 1-\frac{\mu_1}{\mu_2} \\
\frac{H_n-H_{n-k}}{\mu_1}+\frac{H_n-H_{n-1}}{(1-\epsilon)\mu_2}, \ \ otherwise
\end{dcases}
\end{align*}
where $(a)$ holds from Proposition 1 and $(b)$ holds from the Jensen's inequality. Now we prove $(c)$. Let
\begin{equation}
f_{L}=\frac{H_n-H_{n-k+i-1}}{\mu_1}+\frac{H_n-H_{n-i}}{(1-\epsilon)\mu_2}.
\end{equation}
Assuming $n$ is large and $k$ is linear in $n$, we can approximate $H_n\simeq\text{log}(n)$, $H_{n-k}\simeq\text{log}(n-k)$, and $H_{n-k+i-1}\simeq\text{log}(n-k+i-1)$. Then,
\begin{equation}
\frac{\partial^2 f_{L}}{\partial i^2}=\frac{1}{(n-k+i-1)^2\mu_1}+\frac{1}{(1-\epsilon)(n-i)^2\mu_2}>0
\end{equation}
which implies $f_{L}$ is a convex function of $i$. Since $i\in[k]$, the integer $i$ maximizing $f_{L}$ is either $i=1$ or $i=k$. Thus, $(c)$ is satisfied which completes the proof.

\subsection{Upper bound}
For the upper bound, we use the following proposition which is illustrated in Fig. \ref{fig:Proposition2}.
\begin{proposition}
For any two sequences $\{a_i\}_{i\in[n]}$, $\{b_i\}_{i\in[n]}$,
\begin{equation}
\underset{i\in[n]}{k^{\text{th}}\text{min}}(a_i+b_i)\leq\underset{i\in[n]\setminus[k-1]}{\text{min}}(a_{(n+k-i)}+b_{(i)}).
\end{equation}
\end{proposition}
\begin{IEEEproof}
We use the similar idea as the proof of Proposition 1. Consider a set of ordering vectors $\Pi$ defined in (\ref{eq:ordervector}). Note that the $k^{\text{th}}$ smallest value out of $n$ can be always rewritten as the minimum of $n-k+1$ largest values. We define $\boldsymbol{\pi}=[\pi_{1}, ...,\pi_{n}]\in\Pi$ to make the set of values $\{a_{\pi_i}+b_{\pi_i}\}_{i=k}^{n}$ be the $n-k+1$ largest among $\{a_i+b_i\}_{i=1}^{n}$. In addition, define $\boldsymbol{\pi}_1,\boldsymbol{\pi}_2\in\Pi$ to make $a_{\pi_i}=a_{(\pi_{i,1})}$ and $b_{\pi_i}=b_{(\pi_{i,1})}$ be satisfied $\forall i\in[n]\setminus[k-1]$, where $\boldsymbol{\pi}_1=[\pi_{1,1}, ...,\pi_{n,1}]$, $\boldsymbol{\pi}_2=[\pi_{1,2}, ...,\pi_{n,2}]$. Then, $\underset{i\in[n]}{k^{\text{th}}\text{min}}(a_i+b_i)$ can be rewritten as  
\begin{figure}[t]
\centering
     \includegraphics[width=0.46\textwidth, trim=-0.5cm 0 0 0]{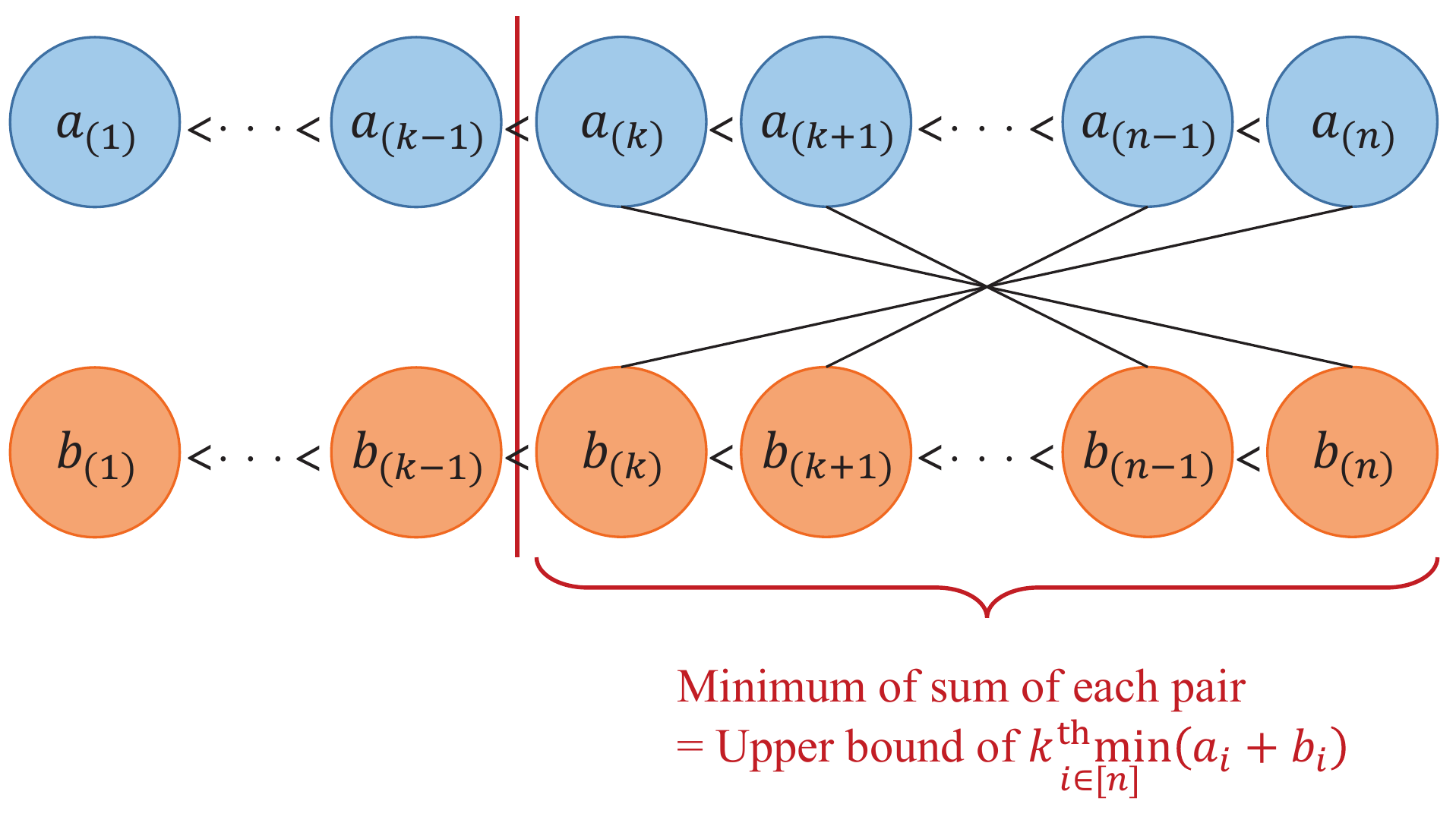}
\caption{Illustration of Proposition 2.}\label{fig:Proposition2}
\end{figure}
\begin{align}
\underset{i\in[n]}{k^{\text{th}}\text{min}}(a_i+b_i)&= \underset{i\in[n]\setminus[k-1]}{\text{min}}(a_{\pi_i}+b_{\pi_i}) \label{eq:kthmin_U1}\\
&= \underset{i\in[n]\setminus[k-1]}{\text{min}}(a_{(\pi_{i,1})}+b_{(\pi_{i,2})}). \label{eq:kthmin_U2}
\end{align}
Now define another set of ordering vectors $\Pi^*\subseteq\Pi$ as $\Pi^*=\{\boldsymbol{\pi}\in\Pi:\pi_i\in[n]\setminus[k-1] \ \forall i\in[n]\setminus[k-1]\}$. Then we can always construct $\boldsymbol{\pi}_1^*, \boldsymbol{\pi}_2^*\in \Pi^*$ such that $\pi_{i,1}\leq\pi_{i,1}^*$, $\pi_{i,2}\leq\pi_{i,2}^*$ for all $i\in [n]\setminus[k-1]$, which can be proved easily. Then by adequately selecting $\boldsymbol{\pi}^*\in \Pi^*$, (\ref{eq:kthmin_U2}) is upper bounded as
\begin{align}
\underset{i\in[n]\setminus[k-1]}{\text{min}}(a_{(\pi_{i,1})}+b_{(\pi_{i,2})})&\leq\underset{i\in[n]\setminus[k-1]}{\text{min}}(a_{(\pi_{i,1}^*)}+b_{(\pi_{i,2}^*)})\label{eq:U_inequal}\\
&=\underset{i\in[n]\setminus[k-1]}{\text{min}}(a_{(\pi_i^*)}+b_{(i)})\label{eq:U_11}
\end{align}
Equation (\ref{eq:U_inequal}) implies that the upper bound of $\underset{i\in[n]\setminus[k-1]}{\text{min}}(a_{(\pi_{i,1})}+b_{(\pi_{i,2})})$ can be obtained by properly selecting the $n-k+1$ pairs $\{a_{(\pi_{i,1}^*)}+b_{(\pi_{i,2}^*)}\}_{i=k}^{n}$ all in the right region of Fig. \ref{fig:Proposition2}, and taking the minimum. This value $\underset{i\in[n]\setminus[k-1]}{\text{min}}(a_{(\pi_{i,1}^*)}+b_{(\pi_{i,2}^*)})$ can be simply rewritten as (\ref{eq:U_11}). Now we want to show that for any $\boldsymbol{\pi}^*\in\Pi^*$, (\ref{eq:U_11}) cannot be larger than $\underset{i\in[n]\setminus[k-1]}{\text{min}}(a_{(n+k-i)}+b_{(i)})$, which is the upper bound illustrated in Fig. \ref{fig:Proposition2}.

Define $U$ as
\begin{equation}
U=\underset{i\in[n]\setminus[k-1]}{\text{min}}(a_{(n+k-i)}+b_{(i)})=a_{(n+k-j)}+b_{(j)}
\end{equation}
for some $j\in[n]\setminus[k-1]$. Suppose there exist $U'$ and an ordering vector $\boldsymbol{\pi'}=[\pi_1', ...,\pi_n']\in\Pi^*$ such that
\begin{equation}\label{eq:contr2}
\underset{i\in[n]\setminus[k-1]}{\text{min}}(a_{(\pi_i')}+b_{(i)})=U'>U.
\end{equation}
We show that the assumption of (\ref{eq:contr2}) is a contradiction. By (\ref{eq:contr2}), $a_{(\pi_{j'}')}+b_{(j')}>a_{(n+k-j)}+b_{(j)}$ should be satisfied for all $j'\in[n]\setminus[k-1]$. This implies $\pi_{j'}'>n+k-j$ for all $j'\in [j]\setminus[k-1]$, which is a contradiction since one-to-one mapping between $j'$ and $\pi_{j'}'$ cannot be constructed for $j'\in[n]\setminus[k-1]$.
\end{IEEEproof}

Now from (\ref{eq:expected_xi}) and (\ref{eq:expected_si}), the upper bound is derived as
\begin{align}
\mathbb{E}[T]&=\mathbb{E}[\underset{i\in[n]}{k^{\text{th}}\text{min}}(X_i+S_i)]\\
&\underset{(d)}{\leq}\mathbb{E}[\underset{i\in[n]\setminus[k-1]}{\text{min}}(X_{(n+k-i)}+S_{(i)})]\\
&\underset{(e)}{\leq}\underset{i\in[n]\setminus[k-1]}{\text{min}}(\frac{H_n-H_{i-k}}{\mu_1}+\frac{H_n-H_{n-i}}{(1-\epsilon)\mu_2})
\end{align}
where $(d)$ holds from Proposition 2 and $(e)$ holds from the Jensen's inequality.

\section{Proof of Theorem 3}
Since $\text{Pr}[X_i\leq t]=1-e^{-\frac{k}{m}\mu_1t}$ for general $k$, we have
\begin{equation}
\mathbb{E}[X_{(i)}]=\frac{m(H_n-H_{n-i})}{k\mu_1} 
\end{equation}
for all $i\in[n]$. For communication time $S_i$, we can first see from Lemma 1 that random variable $\sum_{j=1}^{N_r}Y_{j,r}$ follows exponential distribution with rate $(1-\epsilon)\mu_2$ for all $r\in[\frac{m}{k}]$. Therefore, $S_i=\sum_{r=1}^{\frac{m}{k}}\sum_{j=1}^{N_r}Y_{j,r}$ which is the sum of $\frac{m}{k}$ independent exponential random variables with rate $(1-\epsilon)\mu_2$, follows erlang distribution with shape parameter $\frac{m}{k}$ and rate parameter $(1-\epsilon)\mu_2$:
\begin{equation}
\text{Pr}[S_i\leq t]=1-\sum_{p=0}^{\frac{m}{k}-1}\frac{1}{p!}e^{-(1-\epsilon)\mu_2t}[(1-\epsilon)\mu_2t]^p. 
\end{equation}
By the property of erlang distribution, $S_i$ can be rewritten as 
\begin{equation}
S_i=\frac{G_i}{(1-\epsilon)\mu_2}
\end{equation}
where $G_i$ is an erlang random variable with shape parameter $\frac{m}{k}$ and rate parameter $1$. We take this approach since closed-form expression of $\mathbb{E}[G_{(i)}]$ exists as (\ref{eq:erlangmean}). Now the lower bound is derived as
\begin{align*}
\mathbb{E}[T]&=\mathbb{E}[\underset{i\in[n]}{k^{\text{th}}\text{min}}(X_i+S_i)]\\
&\underset{(a)}{\geq} \mathbb{E}[\underset{i\in[k]}{\text{max}}(X_{(k-i+1)}+S_{(i)})]\\
&\underset{(b)}{\geq} \underset{i\in[k]}{\text{max}}\Big(\frac{m(H_n-H_{n-k+i-1})}{k\mu_1}+\frac{\mathbb{E}[G_{(i)}]}{(1-\epsilon)\mu_2}\Big)\\
\end{align*}
where $(a)$ holds from Proposition 1 and $(b)$ holds from the Jensen's inequality. For the upper bound, we write
\begin{align}
\mathbb{E}[T]&=\mathbb{E}[\underset{i\in[n]}{k^{\text{th}}\text{min}}(X_i+S_i)]\\
&\underset{(d)}{\leq}\mathbb{E}[\underset{i\in[n]\setminus[k-1]}{\text{min}}(X_{(n+k-i)}+S_{(i)})]\\
&\underset{(e)}{\leq}\underset{i\in[n]\setminus[k-1]}{\text{min}}\Big(\frac{m(H_n-H_{i-k})}{k\mu_1}+\frac{\mathbb{E}[G_{(i)}]}{(1-\epsilon)\mu_2}\Big)
\end{align}
where $(d)$ holds from Proposition 2 and $(e)$ holds from the Jensen's inequality.
\ifCLASSOPTIONcaptionsoff
  \newpage
\fi



%


%

\vfill




\end{document}